\documentclass[aps,prl,twocolumn,showpacs,preprintnumbers,amsmath,amssymb]{revtex4-1}
\usepackage[dvipdfmx]{graphicx}
\usepackage{amsmath,braket}
\usepackage{graphicx,float,hyperref,color}

\newcommand{\de}{\partial}
\newcommand{\be}{\begin{equation}}
\newcommand{\ba}{\begin{eqnarray}}
\newcommand{\ea}{\end{eqnarray}}
\newcommand{\ee}{\end{equation}}

\newcommand{\s}{\sqrt}

\newcommand{\ti}{\tilde}

\newcommand{\no}{\nonumber \\}
\newcommand{\la}{\langle}
\newcommand{\lb}{\rangle}
\newcommand{\bea}{\begin{eqnarray}}
\newcommand{\eea}{\end{eqnarray}}
\newcommand{\bes}{\begin{equation*}}
\newcommand{\beas}{\begin{eqnarray*}}
\newcommand{\eeas}{\end{eqnarray*}}
\newcommand{\bas}{\begin{array*}}
\newcommand{\eas}{\end{array*}}
\newcommand{\ees}{\end{equation*}}

\newcommand{\ep}{\epsilon}

\newcommand{\bpm}{\begin{pmatrix}}
\newcommand{\epm}{\end{pmatrix}}
\newcommand{\bbm}{\begin{bmatrix}}
\newcommand{\ebm}{\end{bmatrix}}

\def\CT{{\mathcal{T}}}

\usepackage{bm}

\begin{document}

\begin{flushleft}
YITP-20-148 ; IPMU20-120; MPP-2020-203
\end{flushleft}

\title{Pseudo Entropy in Free Quantum Field Theories}
\author{Ali Mollabashi$^{a,b}$, Noburo Shiba$^b$, Tadashi Takayanagi$^{b,c,d}$, 
Kotaro Tamaoka$^b$, and Zixia Wei$^{b}$}

\affiliation{$^a$Max-Planck-Institut for Physics\\
Werner-Heisenberg-Institut 80805 Munich, Germany}

\affiliation{$^b$Yukawa Institute for Theoretical Physics,
Kyoto University, \\
Kitashirakawa Oiwakecho, Sakyo-ku, Kyoto 606-8502, Japan}

\affiliation{$^c$Inamori Research Institute for Science,\\
620 Suiginya-cho, Shimogyo-ku,
Kyoto 600-8411 Japan}

\affiliation{$^{d}$Kavli Institute for the Physics and Mathematics
 of the Universe,\\
University of Tokyo, Kashiwa, Chiba 277-8582, Japan}

\date{\today}

\begin{abstract}
Pseudo entropy is an interesting quantity with a simple gravity dual, which generalizes entanglement entropy such that it depends on both an initial and a final state. Here we reveal the basic properties of pseudo entropy in quantum field theories by numerically calculating this quantity for a set of two-dimensional free scalar field theories and the Ising spin chain. We extend the Gaussian method for pseudo entropy in free scalar theories with two parameters: mass $m$ and dynamical exponent $z$. This computation finds two novel properties of Pseudo entropy which we conjecture to be universal in field theories, in addition to an area law behavior. One is a saturation behavior and the other one is non-positivity of the difference between pseudo entropy and averaged entanglement entropy. Moreover, our numerical results for the Ising chain imply that pseudo entropy can play a role as a new quantum order parameter which detects whether two states are in the same quantum phase or not. 

\end{abstract}

\maketitle

\section{1. Introduction}

Entanglement entropy in quantum many-body systems plays significant roles in various subjects of theoretical physics, 
such as condensed matter physics \cite{Vidal:2002rm,Kitaev:2005dm,CC04}, 
particle physics \cite{BKLS,Sr,HLW,Casini:2009sr,Nishioka:2018khk} 
and gravitational physics \cite{RTreview,Rev1,Rev2,Rev3}.
In the anti de-Sitter/ conformal field theory (AdS/CFT) correspondence, \cite{Maldacena:1997re}, entanglement entropy is equal to the area of a minimal surface \cite{RT,HRT}. This directly relates geometric structures in quantum many-body systems to those of spacetimes in gravitational theories.

Recently, a new geometric connection between a minimal area surface and a novel quantity, called pseudo entropy, has been found via AdS/CFT \cite{Nakata:2020fjg}. The pseudo entropy is a generalization of entanglement entropy to a transition between the initial state $|\psi_1\lb$ and the final state $|\psi_2\lb$.
First we introduce the transition matrix $\tau^{1|2}$
\ba
\tau^{1|2}=\frac{|\psi_1\lb\la \psi_2|}{\la \psi_2|\psi_1\lb}.
\ea
We divide the total Hilbert space ${\cal H}_{tot}$ 
into two parts $A$ and $B$ as we do so to define entanglement entropy,
i.e. ${\cal H}_{tot}={\cal H}_A\otimes {\cal H}_B$. We introduce the reduced transition matrix 
$\tau^{1|2}_A=\mbox{Tr}_B[\tau^{1|2}]$, by tracing out ${\cal H}_B$. Finally pseudo entropy is defined by
\ba
S(\tau^{1|2}_A)=-\mbox{Tr}[\tau^{1|2}_A\log\tau^{1|2}_A].  \label{PEF}
\ea
Note that when $|\psi_1\lb=|\psi_2\lb$, this quantity is equal to the ordinary entanglement entropy.
Even though this expression (\ref{PEF}) looks like the von-Neumann entropy, this takes complex values in general because $\tau^{1|2}_A$ is no longer hermitian. However, when we construct the initial and final state by a Euclidean path-integral with a real valued action, $S(\tau^{1|2}_A)$ turns out to be positive \cite{Nakata:2020fjg}, which is the case we will focus on in this article.
Moreover, it was found that the pseudo entropy for holographic CFTs can be computed as the areas of 
minimal surfaces in time-dependent Euclidean asymptotically anti de-Sitter (AdS) backgrounds 
\cite{Nakata:2020fjg}. Such a time-dependent Euclidean space is dual to an inner product $\la \psi_2|\psi_1\lb$ via AdS/CFT \cite{Maldacena:1997re}. In addition to the above importance in gravity, 
pseudo entropy has an intriguing interpretation from quantum information viewpoint, as a measure of quantum entanglement for intermediate states between the initial and the final state \cite{Nakata:2020fjg}.
In this letter we would like to pursuit the next obviously important task, namely,
 to uncover basic properties of pseudo entropy in quantum many-body systems, including quantum field theories and condensed matter systems. 

\section{2. Free Scalar Field Theory}

Consider free scalar field theory in two dimensions as our first example.  
We take into account two parameters in the free scalar theory, which are the mass $m$ and the dynamical exponent $z$. At $z=1$, this describes the relativistic scalar field, while for $z>1$, it is called Lifshitz scalar field, which is invariant under the Lifshitz scaling symmetry $t\to\lambda^z t,\ \ x\to \lambda x$ in the $m\to0$ limit.
Its Hamiltonian is written as 
\bea\label{Lifaction}
H=\frac{1}{2}\int dx \left[\pi^2+(\partial_x^z \phi)^2+m^{2z} \phi^2\right],
\eea
where $\phi$ and $\pi$ are the scalar field and its momentum. 

In order to do concrete calculations, we consider its lattice regularization 
\cite{MohammadiMozaffar:2017nri, He:2017wla, MohammadiMozaffar:2017chk, MohammadiMozaffar:2018vmk} given by the Hamiltonian:
\ba\label{eq:LifH}
H\!=\!\sum_{i=1}^{N}\! \left[\frac{\pi_i^2}{2}\!+\!\frac{m^{2z}}{2}\phi_n^2 \!+ \! \frac{1}{2}\left(\sum_{k=0}^z(-1)^{z+k}{{z}\choose{k}}\! \phi_{i-1+k}\right)^2 \! \right].\nonumber
\ea 
where $N$ is the total lattice size. We define $N_A$ to be the lattice size of subsystem $A$. These models are 
straightforwardly generalized to higher dimensions \cite{MohammadiMozaffar:2017nri, MohammadiMozaffar:2017chk}.

It is known that we can calculate the entanglement entropy in free field theories from correlation functions on $A$ when a quantum state is described by a Gaussian wave functional \cite{Cor}. Though for pseudo entropy, we consider a transition matrix instead of a density matrix, we can remarkably extend this Gaussian calculation via an analytic continuation. This makes numerical computations of pseudo entropy possible, playing a major role below.

Two point functions of $\phi$ and $\pi$ consist the $2N_A\times 2N_A$ matrix $\Gamma$
\ba
&& \Gamma=
\begin{pmatrix}
X & R \\ R^T & P
\end{pmatrix},   \label{matg}\\
 \mbox{where}\ \ \  && X_{ij}=\mbox{Tr}[\phi_i\phi_j\tau^{1|2}_A], \ \ P_{ij}=\mbox{Tr}[\pi_i\pi_j\tau^{1|2}_A], \no
&& R_{ij}=\frac{1}{2}\mbox{Tr}[(\phi_i\pi_j+\pi_i\phi_j)\tau^{1|2}_A].
\ea
As opposed to the standard case where $\tau^{1|2}_A$ is given by a hermitian density matrix $\rho_A$, we find that the matrix $R$ takes complex values, though $X$ and $P$ are real symmetric matrices.
Therefore, we consider a complexified symplectic transformation 
$Sp(2N_A,\mathbb{C})$ to diagonalize $\Gamma$ into the form (see appendix A for more details)
\ba
\Gamma\to \begin{pmatrix}
\nu & 0 \\ 0 & \nu
\end{pmatrix},  \label{matgg}
\ea
where $\nu$ is a diagonal matrix and we write its diagonal components as 
$\nu_i=\frac{1}{2}\coth\frac{\ep_i}{2}$. Practically, we can obtain $\nu_i$ from the fact that 
the eigenvalues of the following rearranged matrix are $\pm \nu_i$:
\ba
\begin{pmatrix}
iR^T & iP \\ -iX & -iR
\end{pmatrix}.   \label{matag}
\ea

In our interested examples below, $\nu_i$ and $\ep_i$ always 
take positive real values. Finally, the pseudo entropy is computed by the formula 
\ba
&& S(\tau^{1|2}_A)=\sum_{i=1}^{N_A}\left[\frac{\ep_i}{e^{\ep_i}-1}-\log(1-e^{-\ep_i})\right].\no
&=&\sum_{i=1}^{N_A}\left[\left(\nu_i+\frac{1}{2}\right)\log\left(\nu_i+\frac{1}{2}\right)
-\left(\nu_i-\frac{1}{2}\right)\log\left(\nu_i-\frac{1}{2}\right)\right].\nonumber
\ea
This Gaussian calculation of pseudo entropy can also be justified by a more direct approach, the operator method \cite{Shiba2014, Shiba2020} as presented in appendix B. Though it has not been proven rigorously that performing the analytic continuation used in this Gaussian calculation is possible, we can directly derive the same formula by the operator method without using the analytic continuation.
In our analysis, we take $|\psi_1\lb$ and $|\psi_2\lb$ to be ground states 
for various values of the mass $m$ and dynamical exponent $z$, which we denote by $(m_1,z_1)$ and $(m_2,z_2)$.

\begin{figure}[h!]
\includegraphics[scale=.5]{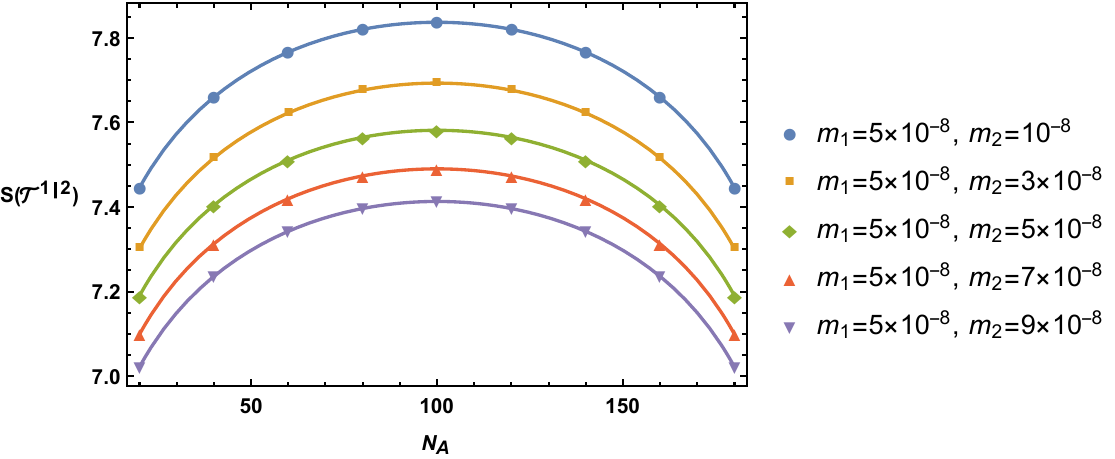}
\caption{$S(\tau^{1|2}_A)$ as a function of the size of the subsytem $N_A$. We set $N=200$ and $z_1=z_2=1$. The curves are $c_1 \ln[(N/\pi)\sin[\pi N_A/N]]+c_0$, where $c_1 \simeq 0.3333$ and $6.028<c_0 <6.453$.
}
\label{GraphPEvsSizeN200Finite01}
\end{figure}

Let us first start with the relativistic setups $z_1=z_2=1$ and $m_1\neq m_2$. We take the total system to be  a circle length $L$ and define a subsystem $A$ to be a length $l$ interval on this circle.  We write the UV cut off (lattice spacing) as $\ep$ such that $L=N\ep$ and $l=N_A\ep$.
Our numerical analysis reveals the general behavior of pseudo entropy
\ba
S(\tau^{1|2}_A)=\frac{1}{3}\log\left(\frac{L}{\pi\ep}\sin\left(\frac{\pi l}{L}\right)\right)+f(m_1,m_2,L,l),
\label{staul}
\ea
where the first term in the right hand side coincides with the known behavior of entanglement entropy in two dimensional CFT with the central charge $c=1$ \cite{HLW,CC04}, while the second term is a constant term which depends on the relevant parameters. For confirmations of this behavior,  refer to Fig.\ref{GraphPEvsSizeN200Finite01}, where the first logarithmic term in (\ref{staul})  gives a dominant $l$ dependence for small masses.  This shows that the leading logarithmic divergence, which is equivalent to the area law, is robust for the pseudo entropy.
For small values of masses, our numerical calculations determine analytical structures of the function $f(m_1,m_2,L,l)$. When we consider the almost massless limit $m_{1,2}L\ll 1$, we have 
\ba
f(m_1,m_2,L,l)\simeq -\frac{1}{2}\log\left[\frac{m_1+m_2}{2}L\right], \label{LMAS}
\ea
as we will explain in appendix C. This logarithmic behavior is due to the zero mode of scalar field and the above formula agrees with the known result of entanglement entropy in \cite{Casini:2004bw}.
When the mass is small such that $m_{1,2}L\sim 1$ and $m_{1,2}l\ll 1$, we can find the $l$ dependence 
\ba
&& f(m_1,m_2,L,l)  \no
&& \simeq\dfrac{1}{2}\log\left[-\dfrac{m_1^2\log[m_1 l]-m_2^2\log[m_2 l]}{m_1^2-m_2^2}\right]+f_0(m_1,m_2,L),\nonumber
\ea
where the final term $f_0$ does not depend on $l$. This expression again reproduces the known 
$\frac{1}{2}\log\left(-\log (ml)\right)$ term \cite{Casini:2005zv} in the entanglement entropy. 
Refer to appendix D for more details.

\begin{figure}[h!]
\begin{center}
\includegraphics[scale=0.35]{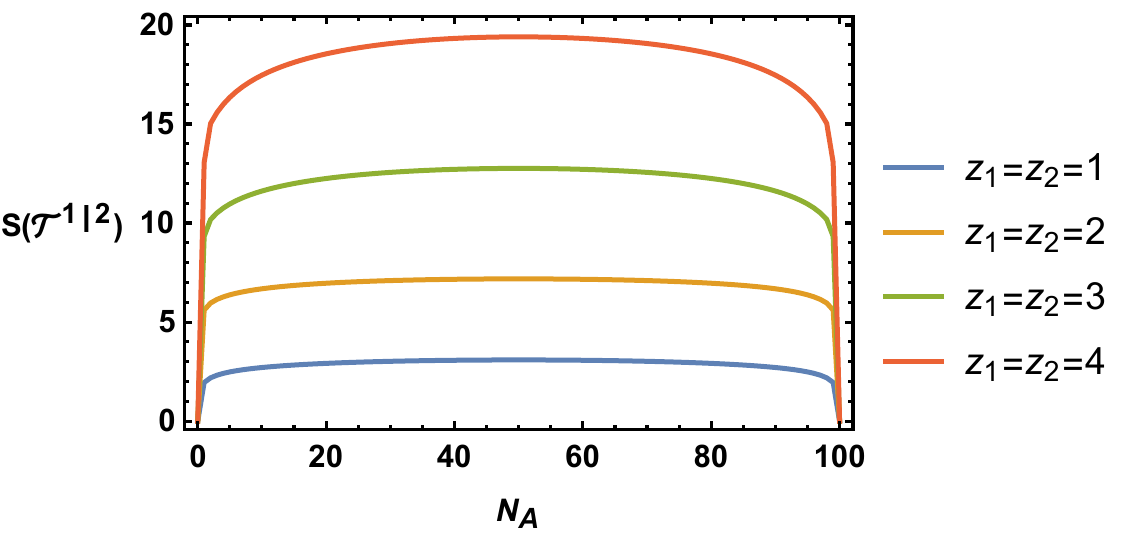}
\includegraphics[scale=0.35]{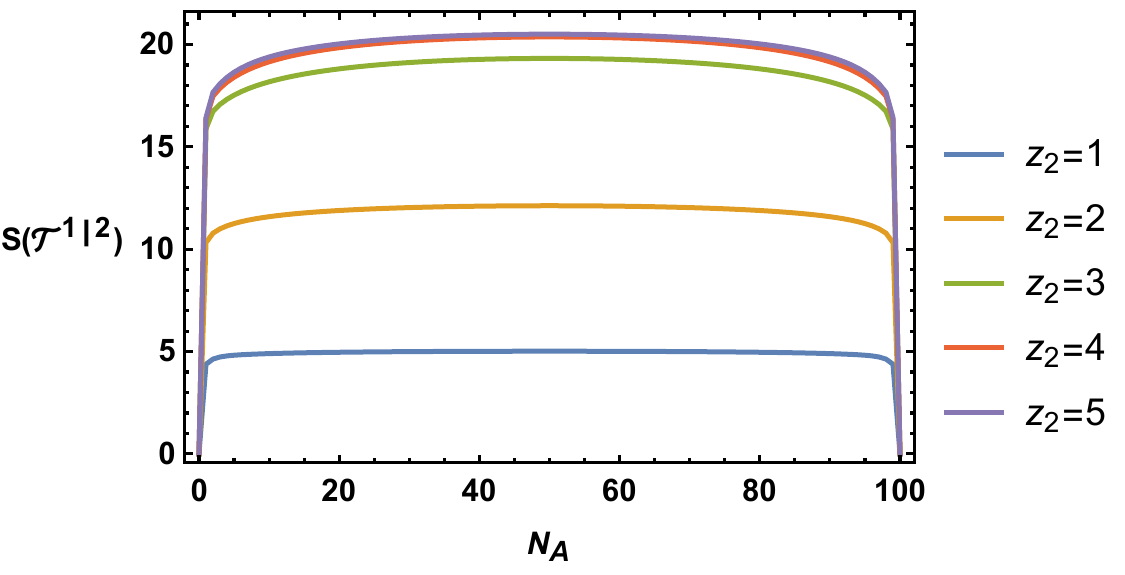}
\end{center}
\caption{
The upper plot shows the pseudo entropy as a function of the subsystem size $N_A$ when we chose  $m_1=10^{-3}$ and $m_2=10^{-5}$
for various values of $z_1=z_2$. The lower plot shows the pseudo entropy when we set $z_1=3$ and $m_1=m_2=10^{-5}$. We chose the total system $N=100$.
}
\label{fig:ccresultas}
\end{figure}

\begin{figure}[h!]
\begin{center}
\includegraphics[scale=0.35]{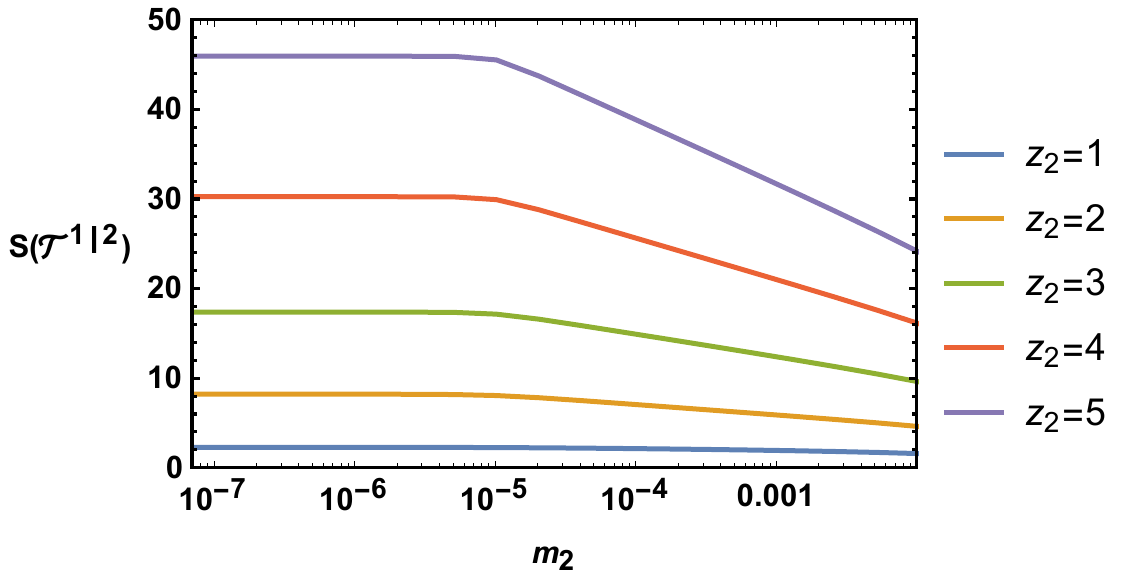}
\includegraphics[scale=0.35]{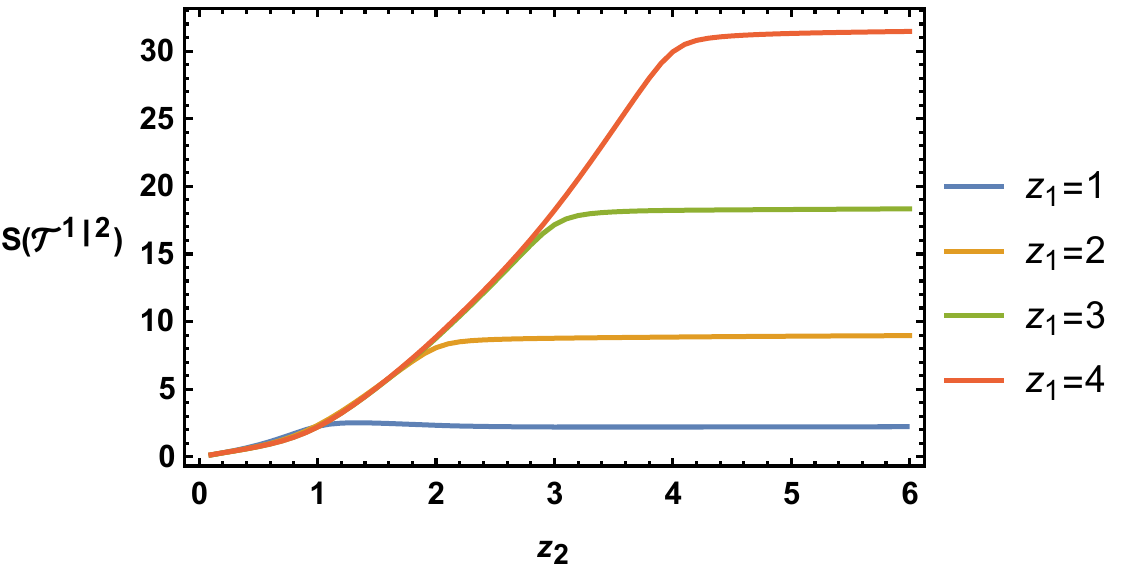}
\end{center}
\caption{The upper graph shows the pseudo entropy as a function of $m_2$ when we set $z_1=1$ and $m_1=10^{-5}$. The lower graph depicts  the pseudo entropy as a function of $z_2$ when we set
 $m_1=m_2=10^{-5}$.We chose $N_A=50$ and $N=\infty$. }
\label{fig:abc}
\end{figure}

Now we turn on the dynamical exponent $(z_1,z_2)$ to describe the Lifshitz scalar theory. 
When $z_1=z_2$, the pseudo entropy gets larger as the dynamical exponent increases as in the upper graph of Fig.\ref{fig:ccresultas}. When we fix $z_1$ and increases $z_2$, the pseudo entropy approaches to a certain finite value as can be seen from the lower graph in Fig.\ref{fig:ccresultas}. We call this phenomenon saturation. The saturation occurs when we fix $|\psi_1\lb$ and consider a limit where the entanglement of $|\psi_2\lb$ gets larger. The two graphs in Fig.\ref{fig:abc} demonstrate the saturations when we take different two limits of 
$m_2\to 0$ and $z_2\to\infty$, respectively. This saturation in our free scalar field theory implies that the behavior of pseudo entropy qualitatively looks like 
\ba
S(\tau^{1|2}_A)\sim \mbox{Min} [S(\rho^1_A),S(\rho^2_A)].  \label{minrel}
\ea

From our numerical results, we can find one more basic property of pseudo entropy by introducing the difference:
\ba
\Delta S_{12}\equiv S(\tau^{1|2}_A)-\frac{S(\rho^1_A)+S(\rho^2_A)}{2}.  \label{deltas}
\ea
If $|\psi_1\lb$ and $|\psi_2\lb$ are very closed to a state $|\psi_0\lb$,
such that $\delta \tau_A=\tau^{1|2}_A-\rho^0_A$ is very small, 
then we can derive a first law like relation (see appendix E for a derivation):
\ba
S(\tau^{1|2}_A)-S(\rho^0_A)\simeq \frac{\la \psi_2|H_A|\psi_1\lb}{\la \psi_2|\psi_1\lb}+O\left((\delta \tau_A)^2\right),  \label{firstl}
\ea
as in the first law of entanglement entropy \cite{Bhattacharya:2012mi,Blanco:2013joa,Wong:2013gua}.
 Here we introduced the 
modular Hamiltonian $H_A=-\log\rho^0_A-S(\rho^0_A)$.
 The linear combination (\ref{deltas}) is special such that it cancels out in this
linear difference (\ref{firstl}), leaving only the quadratic order 
as $\Delta S_{12}=O\left((\delta \tau_A)^2\right)$.

\begin{figure}[h!]
\begin{center}
\includegraphics[scale=0.35]{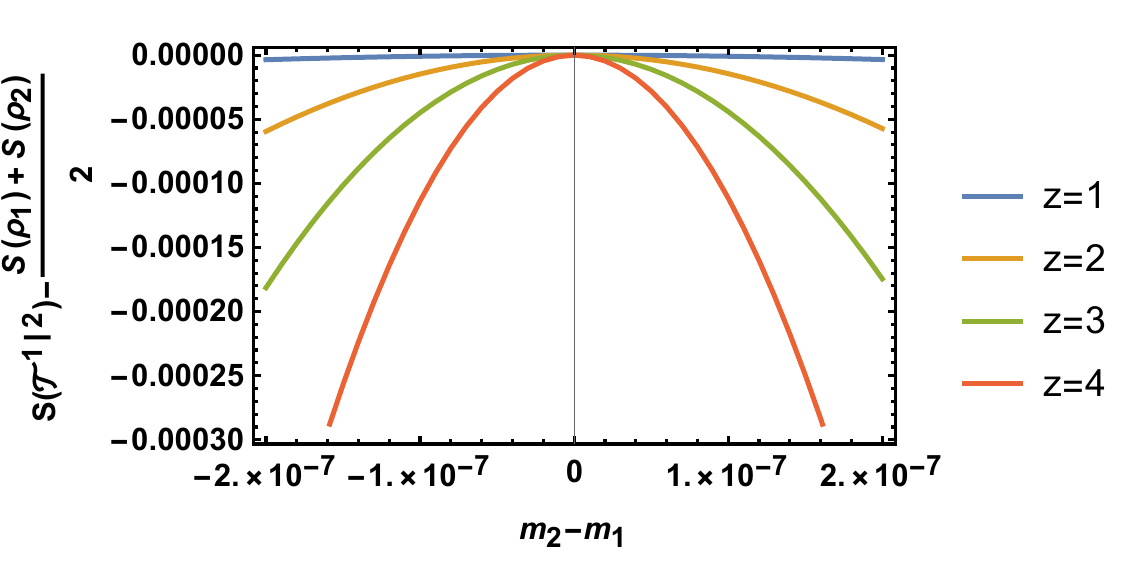}
\includegraphics[scale=0.35]{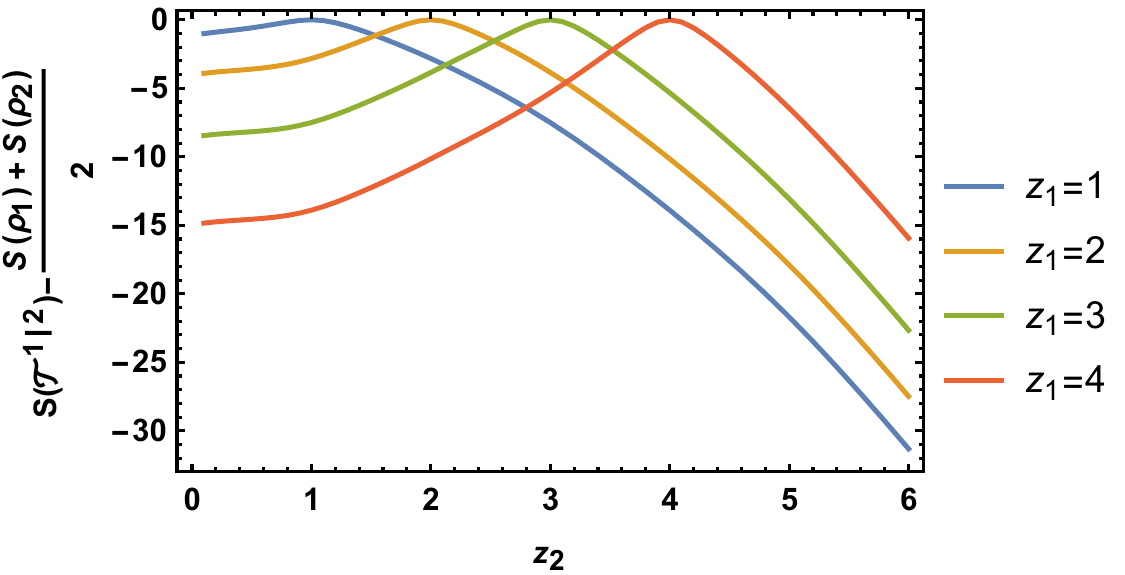}
\end{center}
\caption{The plots of the difference $\Delta S_{12}$ as a function of 
$m_2-m_1$ (upper) and $z_2$ (lower). We set $m_1=10^{-5}$ and $z_1=z_2$ in the upper graph. 
We chose $m_1=m_2=10^{-5}$ in the lower.
}
 \label{fig:ccdrresults}
\end{figure}

In general, this quadratic difference $\Delta S_{12}$ is not guaranteed to be positive definite. Indeed, we can confirm that both signs are possible even in a two qubit example, as discussed in appendix E.
However, in all of our numerical results in the free scalar field theory (\ref{Lifaction}), 
we observe its non-positivity $\Delta S_{12}\leq 0$ 
when we vary the masses and dynamical exponents, as depicted in Fig.\ref{fig:ccdrresults}.
Also, in the small mass limit (\ref{LMAS}), this non-positivity is satisfied.

\section{Pseudo Entropy in Perturbed CFT}

To investigate the behavior of pseudo entropy more, consider a perturbation in a two dimensional CFT.
We assume that the subsystem $A$ is a length $l$ interval and the CFT is defined on R$^2$. The perturbation is expressed as $\lambda \int dtdx O(t,x)$, where $O$ is a primary operator and $\lambda$ is a small perturbation parameter. We choose $|\psi_1\lb$ is the original CFT vacuum and $|\psi_2\lb$ is the new vacuum obtained by this perturbation. Since one point function vanishes in a CFT, there is no $O(\lambda)$ term in the differences $S(\tau^{1|2}_A)-S(\rho^1_A)$. Moreover, at the order  $O(\lambda^2)$, we can show
 $S(\tau^{1|2}_A)-S(\rho^1_A)\leq 0$ as we give the details in appendix F. This result is universal because it only involves two point functions in a CFT.  

In particular, if we consider an exactly marginal perturbation, we find that the coefficient of the logarithmically divergent terms is changed
\ba
S(\tau^{1|2}_A)=\frac{c}{3}f(\lambda)\log\frac{l}{\ep}+\mbox{const.},  \label{stauc}
\ea
The conformal perturbation shows $f(\lambda)=1+g\lambda^2+O(\lambda^3)$ with $g<0$ in the
$\lambda\to 0$ limit. We can also derive the same behavior from the holographic calculation of pseudo entropy in Janus solutions \cite{Bak:2003jk, Freedman:2003ax, Clark:2004sb, Clark:2005te,DHoker:2006vfr, Bak:2007jm}. In this way, we can confirm  $\Delta S_{12}\leq 0$ for exactly marginal perturbations. Refer to appendix F for derivations of these results.

\section{Pseudo Entropy in Ising Model}

As another class of basic quantum many-body systems, we would like to consider a transverse Ising spin chain model. In the continuum limit near the critical point, this model is known to be equivalent to the two dimensional free fermion CFT \cite{Sac}. 
Its Hamiltonian can be written as 
\begin{align}
    H = -J\sum_{i=0}^{N-1} \sigma^z_{i}\sigma^z_{i+1} - h \sum_{i=0}^{N-1} \sigma^x_{i},
\end{align}
where the spins are be labeled by $i=0,1,2,\cdots,N-1$ and the $\sigma_i^z$ is Pauli operator on $i$ with eigenvalues $\pm1$. We impose the periodic boundary condition. Note that the quantum critical point
is situated at $J=h$ in the continuum limit, where  $J>h$ is the ferromagnetic phase, while 
$J<h$ describes the paramagnetic phase.

\begin{figure}[H]
    \centering
    \includegraphics[width=4.2cm]{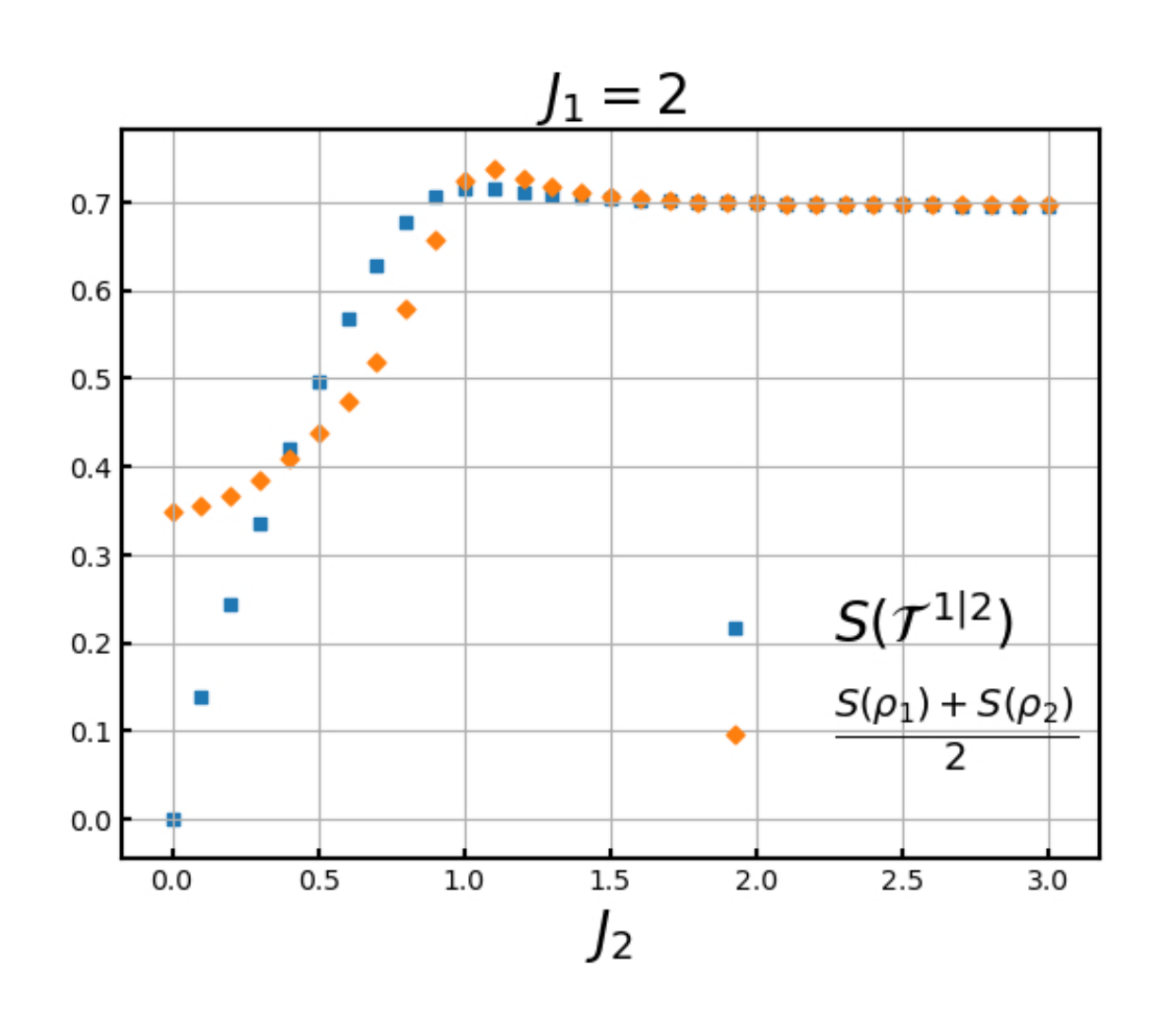}
    \includegraphics[width=4.2cm]{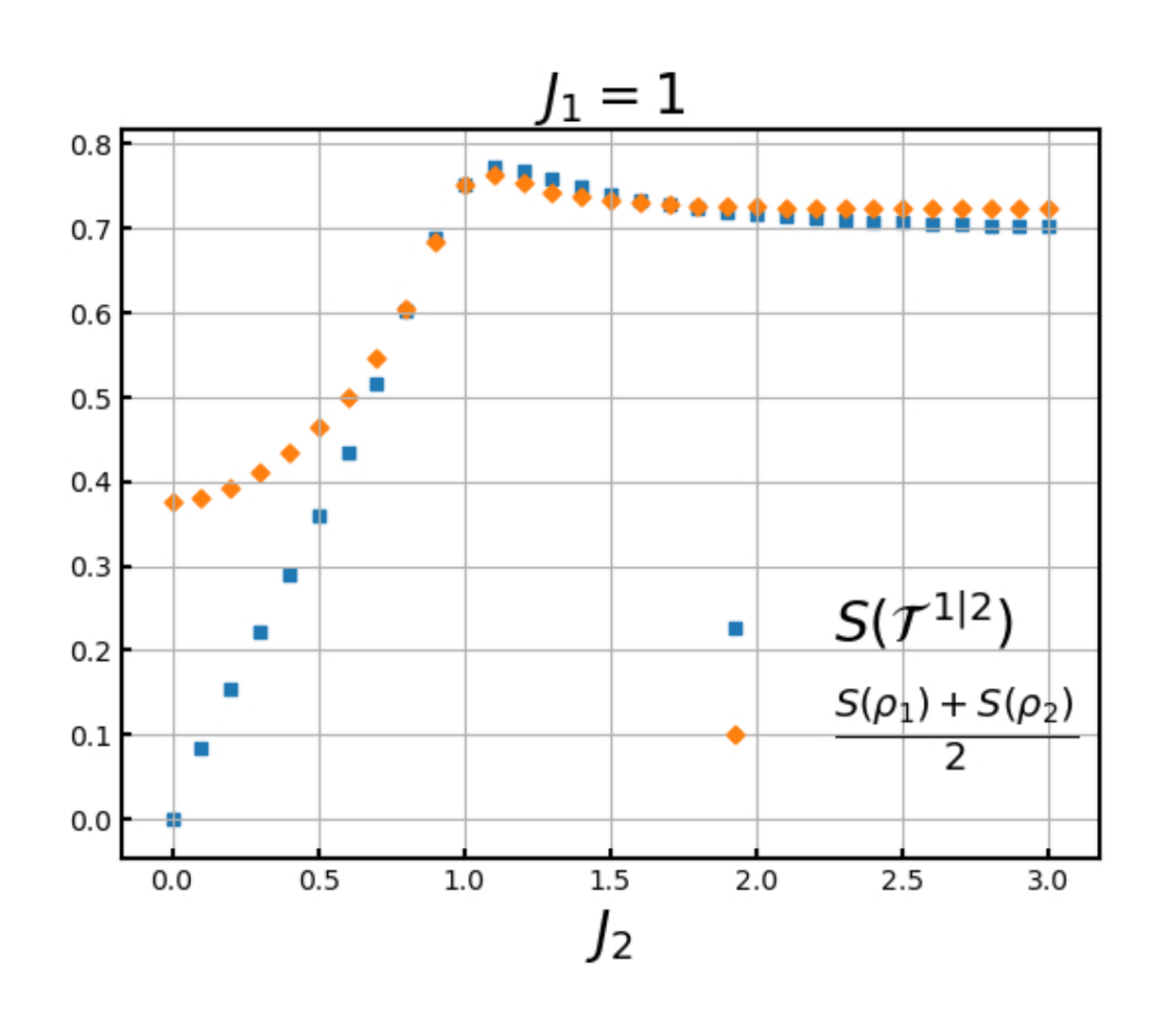}
    \includegraphics[width=4.2cm]{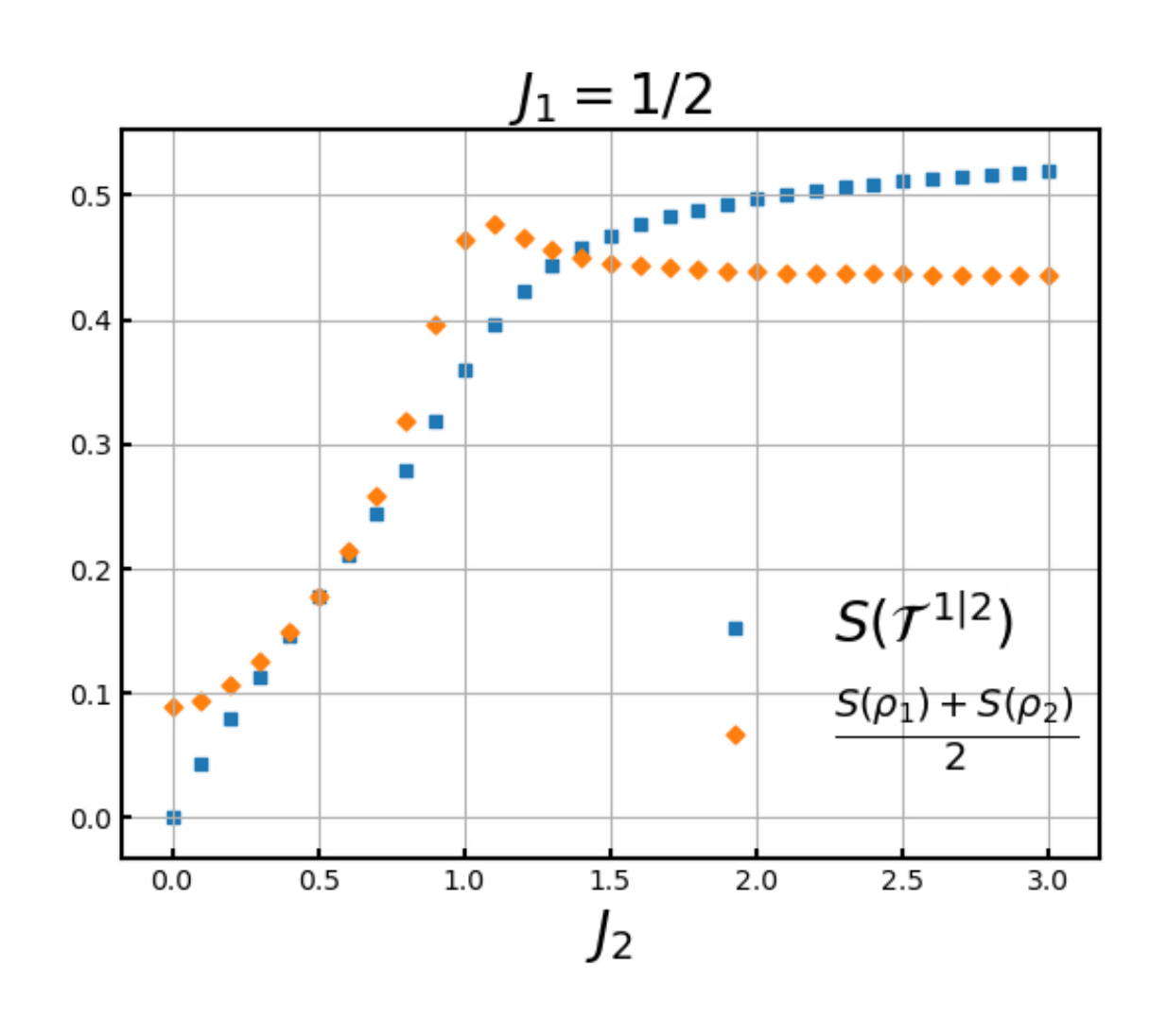}
    \caption{Pseudo entropy and average of entanglement for a single interval. Here, we choose $N=16,N_A=8,~h_1=h_2=1$. We set $J_1=2$ (upper left), $J_1=1$ (upper right) and 
$J_1=1/2$ (lower). The horizontal axis is the value of $J_2$. The blue dots show the pseudo entropy $S(\CT^{1|2}_A)$ and the orange dots show the average of the entanglement entropy $\left(S(\rho^1_A) + S(\rho^2_A)\right)/2$.}
\label{fig:spin}
\end{figure}

We calculate the pseudo entropy $S(\tau^{1|2}_A)$ by choosing $|\psi_1\lb$ and $|\psi_2\lb$ to be the ground states for $(J,h)=(J_1,h_1)$ and $(J_2,h_2)$, respectively. The subsystem $A$ is assumed to be a single interval with $N_A$ spins. We show numerical results in Fig.(\ref{fig:spin}) (We used the python package quspin \cite{WB17} in our computation.).

From the numerical results, we can observe the saturation $S(\tau^{1|2}_A)\simeq \log 2$ in the $J_2\to \infty$ limit when $J_1 >1$. Moreover we can confirm that the difference (\ref{deltas}) satisfies $\Delta S_{12}\leq 0$ when
$(J_1,h_1)$ and  $(J_2,h_2)$ are in the same phase, i.e. $(J_1-h_1)(J_2-h_2)>0$.
However, we can have $\Delta S_{12}>0$ when they belong to two different phases, i.e. $(J_1-h_1)(J_2-h_2)<0$. This implies that the sign of   the difference $\Delta S_{12}$ can provide an order parameter which tells us whether the two states $|\psi_1\lb$ and  $|\psi_2\lb$ are in the same phase or not. 
This result is also expected to hold when considering two ground states of 2D free Majorana fermion theories with different mass as long as they belong to the same phase \cite{longVersion}, since free Majorana fermion can be obtained as a scaling limit of transverse Ising chain after Jordan-Wigner transformations.

\section{Discussions}

In this article we have uncovered basic properties of pseudo entropy in quantum field theories by focusing on numerical calculations in a class of free scalar field theories and the Ising spin chain. 
We would like to conjecture that the properties: area law, saturation and non-positivity of $\Delta S_{12}$, which we found for free scalar field theories, will be universal also for any quantum field theory. It will be an important future problem to study pseudo entropy in broader class of field theories and test the above properties.
Moreover, our results for Ising spin chain imply that we can classify different phases in quantum many-body systems from the calculations of pseudo entropy. This origins from our expectation that the pseudo entropy helps us to probe the difference of structures of quantum entanglement between two states.  
One obvious future direction will be to analyze the pseudo entropy in topological phases, to see if it can play a role of topological order parameter.

\vspace{5mm}
{\bf Acknowledgements} 
We are grateful to Seishiro Ono, Hong Yang and Chi Zhang for useful discussions and to Yoshifumi Nakata, Tatsuma Nishioka, 
and Yusuke Taki for communications on this article.
TT is supported by Grant-in-Aid for JSPS Fellows No.~19F19813.
KT and TT are supported by the Simons Foundation through the ``It from Qubit'' collaboration.  
TT is supported by Inamori Research Institute for Science and 
World Premier International Research Center Initiative (WPI Initiative) 
from the Japan Ministry of Education, Culture, Sports, Science and Technology (MEXT). 
NS and TT are supported by JSPS Grant-in-Aid for Scientific Research (A) No.~16H02182. 
AM and TT are also supported by JSPS Grant-in-Aid for Challenging Research (Exploratory) 18K18766.
KT is also supported by JSPS Grant-in-Aid for Research Activity start-up 19K23441.
AM is generously supported by Alexander von Humboldt foundation via a postdoctoral fellowship.
NS is also supported by JSPS KAKENHI Grant Number JP19K14721. ZW is supported by the ANRI Fellowship and Grant-in-Aid for JSPS Fellows No.20J23116.

\newpage
\widetext
\setcounter{equation}{0}
\setcounter{figure}{0}
\setcounter{table}{0}
\makeatletter
\appendix

\section{Appendix A: Correlation Function Method for Pseudo Entropy}
\subsection{Lifshitz Scalar Theories}
We consider the following free scalar theories,
\ba\label{action}
S=\frac{1}{2}\int dt d\vec{x} \left[\dot{\phi}^2-\sum_{i=1}^{d}(\partial_i^z \phi)^2-m^{2z} \phi^2\right],
\ea
which are invariant under Lifshitz scaling symmetry in the massless limit $(m=0)$. 

In order to do concrete calculations we consider the regularized version of these theories on a lattice, known as Lifshitz harmonic lattice models,
\be\label{eq:LifHa}
H=\sum_{n=1}^{N}\left[\frac{\pi_n^2}{2M}+\frac{M m^{2z}}{2}\phi_n^2+\frac{K}{2}\sum_{i=1}^{d}\left(\sum_{k_i=0}^z(-1)^{z+k_i}{{z}\choose{k_i}} \phi_{n-1+k_i}\right)^2 \right],
\ee 
where we set $M=K=1$ without loss of generality (see \cite{MohammadiMozaffar:2017nri, He:2017wla, MohammadiMozaffar:2017chk, MohammadiMozaffar:2018vmk} where different information theoretic properties of these models has been addressed). The $z=1$ case is the standard harmonic lattice model. The diagonalized Hamiltonian in generic dimensions takes the following form
\bea
H=\sum_{\mathbf{k}}\omega_{\mathbf{k}}\left(a_{\mathbf{k}}a^\dagger_{\mathbf{k}}+\frac{1}{2}\right),
\eea
where
\bea\label{dispersion}
\omega^2_{\mathbf{k}}=m^{2z}+\sum_{i=1}^d\left(2\sin\frac{\pi k_{i}}{N_{x_i}}\right)^{2z}.
\eea
In the following we explain how to compute pseudo entropy in these theories, though the method is more general for ant Gaussian state in quadratic theories. 

\section{Pseudo Entropy in Scalar Theories: Correlator Method}
Standard correlator method is used to study entanglement and Renyi entropies is Gaussian states of quadratic theories. The idea is based on the fact that the spectrum of the reduced density matrix is fully determined with the two-point functions of the operators restricted into the subregion of interest. The idea is very similar in case of pseudo entropy, except that the notion of density matrix is replaced by the transition matrix. The transition matrix in the post-selection setup defines an analogue to the expectation value of these restricted operators on a Gaussian state as
\be\label{eq:trace}
\langle \mathcal{O} \rangle = \frac{\langle \psi_2 | \mathcal{O} | \psi_1 \rangle}{\langle \psi_2 |\psi_1 \rangle}=\mathrm{Tr}
\left[\tau^{1|2}\, \mathcal{O}\right].
\ee
We consider the case when $|\psi_{1,2}\rangle$ are vacuum states with different $(m,z)$ parameters in the Hamiltonian, namely with different dispersion relations.
In this case we have $a_{1}|\psi_1\rangle=a_{2}|\psi_2\rangle=0$. As will be described in the appendix B with more detail, these states are related to each other via 
\begin{align}
\begin{split}
a_{2}&=\alpha_k\, a_{1} + \beta_k\, a^{\dagger}_{1}\,,
\\
a^{\dagger}_{2}&=\beta_k\, a_{1} + \alpha_k\, a^{\dagger}_{1}\,,
\end{split}
\end{align}
where
\be
\alpha_k=\frac{1}{2}\left(\sqrt{\frac{\omega_k^{(2)}}{\omega_k^{(1)}}}+\sqrt{\frac{\omega_k^{(1)}}{\omega_k^{(2)}}}\right)
\;\;\;\;\;\;
,
\;\;\;\;\;\;
\beta_k=\frac{1}{2}\left(\sqrt{\frac{\omega_k^{(2)}}{\omega_k^{(1)}}}-\sqrt{\frac{\omega_k^{(1)}}{\omega_k^{(2)}}}\right),
\ee
and $\omega^{(i)}$'s are determined by $(m_i,z_i)$ in \eqref{dispersion}.  

With the above Bogoluibov transformations, we can determine $|\psi_{2}\rangle$ in terms of the eigenvectors of the number operator $n_{1}=a^{\dagger}_{1}a_{1}$ as
\be
|\psi_2 \rangle=\sum_{n=0}^\infty c_{2n}\,| 2n\rangle_{1}
\;\;\;\;\;\;\;\;,\;\;\;\;\;\;\;\;
c_{2n}=\left(-\frac{\beta}{\alpha}\right)^n\sqrt{\frac{(2n-1)!!}{2n!!}}\;c_0
\ee
The expectation values of the restricted operators in two dimensions on a translational invariant lattice are given by 
\begin{align}
\begin{split}
X_{rs}&\equiv
\frac{\langle \psi_2 | \phi_r\phi_s | \psi_1 \rangle}{\langle \psi_2 |\psi_1 \rangle}
=
\frac{1}{2N}\sum_{k=0}^{N-1} \frac{2}{\omega_k^{(2)}+\omega_k^{(1)}}\cos\left(\frac{2\pi k (r-s)}{N}\right)
\\
P_{rs}&\equiv
 \frac{\langle \psi_2 | \pi_r\pi_s | \psi_1 \rangle}{\langle \psi_2 |\psi_1 \rangle}
=
\frac{1}{2N}\sum_{k=0}^{N-1} \frac{2\omega_k^{(1)}\omega_k^{(2)}}{\omega_k^{(2)}+\omega_k^{(1)}}\cos\left(\frac{2\pi k (r-s)}{N}\right)
\\
R_{rs}&\equiv
\frac{1}{2} \frac{\langle \varphi | (\phi_r\pi_s+\pi_r\phi_s) | \psi \rangle}{\langle \varphi |\psi \rangle}
=
\frac{i}{2N}\sum_{k=0}^{N-1} \frac{\omega_k^{(2)}-\omega_k^{(1)}}{\omega_k^{(2)}+\omega_k^{(1)}}\cos\left(\frac{2\pi k (r-s)}{N}\right)
\end{split}
\end{align}
where $r,s=1,2,\cdots,N_A$. In this case as opposed to entanglement and Renyi entropies in static states the $R$ correlators, which take \textit{pure imaginary} values in our case, play a non-trivial role. 
In order to find a suitable transformation that brings the transition matrix to a diagonal form
\be\label{eq:diagT}
\tau^{1|2}=\bigotimes_{k=1}^{N_A}\left(1-e^{-\epsilon_k}\right)e^{-\epsilon_k n_{A,k}},
\ee
we need a transformation which preserves the commutation relations. To this end we consider a generalized vector of canonical variables, the fields and their conjugate momenta, as $r=\left( \phi_1,\cdots,\phi_{N_A},\pi_1,\cdots,\pi_{N_A} \right)^T$. So the canonical commutation relations read
\be
[r_\alpha,r_\beta]=i J_{\alpha\beta}
\;\;\;\;\;\;\;,\;\;\;\;\;\;\;
J=
\begin{pmatrix}
0 & \mathbf{1} \\ -\mathbf{1} & 0
\end{pmatrix},
\ee
where $A,B=1,2,\cdots,2N$, and we define a correlator matrix as
\be
\Gamma_{\alpha\beta}
\equiv
\frac{1}{2}\langle \{r_\alpha,r_\beta\} \rangle
=
\begin{pmatrix}
X_{ab} & R_{ab} \\ R_{ab}^T & P_{ab}
\end{pmatrix}
\ee
We consider the following transformations between the creation and annihilation operators restricted to subregion $A$ 
\be
\phi_r = \alpha^*_{rk} a_{A,k}^\dagger+ \alpha_{rk} a_{A,k}, 
\;\;\;\;\;\;,\;\;\;\;\;\;
\pi_r = -i\,\beta^*_{rk} a_{A,k}^\dagger + i\,\beta_{rk}a_{A,k}\;,
\ee
where from commutation relations we find
\be
\alpha^*\cdot\beta^T+\alpha\cdot\beta^\dagger=-1
\;\;\;\;\;\;,\;\;\;\;\;\;
\alpha^*\cdot\alpha^T=\alpha\cdot\alpha^\dagger
\;\;\;\;\;\;,\;\;\;\;\;\;
\beta^*\cdot\beta^T=\beta\cdot\beta^\dagger\;\;\;.
\ee
These transformations lead to the following expressions for the correlators
\begin{align}
\begin{split}
X&=\alpha^*\cdot\nu\cdot\alpha^T+\alpha\cdot\nu\cdot\alpha^\dagger,
\\
P&=\beta^*\cdot\nu\cdot\beta^T+\beta\cdot\nu\cdot\beta^\dagger,
\\
R&=i\left(\alpha^*\cdot\nu\cdot\beta^T-\alpha\cdot\nu\cdot\beta^\dagger\right).
\end{split}
\end{align}
In case of dealing with density matrices, where the $R$ correlators take real values, utilizing Williamson's theorem \cite{Williamson}, for any symmetric positive definite $\Gamma$ there always exists a symplectic transformation, $S\in$ Sp($2N,\mathbb{R}$) such that
\be  
r'=S\cdot r
\;\;\;\;\;\;\;
,
\;\;\;\;\;\;\;
J=S\cdot J\cdot S^T
\;\;\;\;\;\;\;
,
\;\;\;\;\;\;\;
\Gamma
=
\begin{pmatrix}
\mathrm{diag}(\nu_k) & \mathbf{0} \\ \mathbf{0} & \mathrm{diag}(\nu_k)
\end{pmatrix}\;.
\ee
Now that the $R$ correlator is pure imaginary, although the original form of Williamson's theorem does not apply, we consider an analytic continuation of such a transformation, i.e. $S\in$ Sp($2N,\mathbb{C}$) 
\cite{ANW}. This continuation is non-singular in our criteria of interest, as we provide several justifications in the following appendix as well as in \cite{longVersion}. An easy way to work out $\{\epsilon_k\}$ is to find the spectrum of $(i J\cdot\Gamma)$ denoted by $\{\nu_k\}$, which gives a double copy of $\{\epsilon_k\}$ as
\be
\nu_k=\pm \frac{1}{2}\coth\left(\frac{\epsilon_k}{2}\right)
\ee
In the following appendix B, alternatively we use the operator method to directly prove that even without assuming any ansatz for the transition matrix \eqref{eq:diagT}, pseudo entropy can be directly read from the spectrum of $(i J\cdot\Gamma)$.

\section{Appendix B: Operator method for Pseudo Entropy}\label{ap:OP}
We calculate the pseudo entropy by using the operator method developed in \cite{Shiba2014,Shiba2020}.
First, we summarize the Bogoliubov transformation. Next, we calculate the pseudo entropy.

\subsection{Bogoliubov transformation}\label{subsection BT}
We consider a real free scalar field in $(d+1)$ dimensional spacetime. 
As an ultraviolet regulator, we replace the continuous $d$-dimensional space coordinates $x$ by a lattice of discrete points with spacing $a$. 
As an infrared cutoff, we allow the individual components of $n \equiv x/a$ to assume only a finite number $N$ of independent values $-N/2<n_{\mu} \leq N/2 .$
The Greek indices denoting vector quantities run from one to $d$.
Outside this range we assume the lattice is periodic. 
The scalar field $\phi_{n}$ and the conjugate momentum $\pi_{n}$ obey the canonical commutation relations 
\begin{equation}
[\phi_{n} , \pi_{m}]=i \delta_{nm}  . \label{eq:4-3}
\end{equation}
We consider vacuum states $\ket{0}_{\alpha}~~(\alpha=1,2)$ of Hamiltonians $H_{\alpha}$, 
\begin{equation}
H_{\alpha}=\sum_{k} \omega_k^{(\alpha)} a_k^{(\alpha) \dagger} a_k^{(\alpha )},
\end{equation}
where the index $k$ also carries $d$ integer valued components, each in the range of $-N/2<k_{\mu} \leq N/2 $ and $\left[a_k^{(\alpha )}, a_{k'}^{(\alpha )\dagger}\right]=\delta_{k, k'}$ and $\omega_k^{(\alpha)}=\omega_{-k}^{(\alpha)}$. 
We expand $\phi_{n}$ and $\pi_{n}$ as 
\begin{equation}
\begin{split}
&\phi_{n} = \frac{1}{N^{d/2}} \sum_{k} \frac{1}{\sqrt{2\omega_k^{(\alpha)}}} 
\left[e^{2\pi i kn/N} a_k^{(\alpha )} + e^{-2\pi i kn/N} a_k^{(\alpha )\dagger} \right], \\
&\pi_{n} = \frac{1}{N^{d/2}} \sum_{k} (-i) \sqrt{ \frac{\omega_k^{(\alpha)}}{2} }
\left[e^{2\pi i kn/N} a_k^{(\alpha )} - e^{-2\pi i kn/N} a_k^{(\alpha )\dagger} \right].
\end{split}  \label{fourier}
\end{equation}
From (\ref{fourier}), we obtain
\begin{equation}
\begin{split}
&a_k^{(\alpha )} = \sqrt{ \frac{\omega_k^{(\alpha)}}{2} } \tilde{\phi}_{k} + \frac{i}{\sqrt{2\omega_k^{(\alpha)}}}  \tilde{\pi}_{k}  \\
&a_{-k}^{(\alpha )\dagger} = \sqrt{ \frac{\omega_k^{(\alpha)}}{2} } \tilde{\phi}_{k} - \frac{i}{\sqrt{2\omega_k^{(\alpha)}}}  \tilde{\pi}_{k}  ,
\end{split}  \label{annihilation}
\end{equation}
where
\begin{equation}
\begin{split}
&\tilde{\phi}_{k}\equiv \frac{1}{N^{d/2}} \sum_{k} e^{-2\pi i kn/N} \phi_{n} , \\
&\tilde{\pi}_{k}\equiv \frac{1}{N^{d/2}} \sum_{k} e^{-2\pi i kn/N} \pi_{n} .
\end{split}  \label{fourier inverse}
\end{equation}
From (\ref{annihilation}), we obtain the Bogoliubov transformation, 
\begin{equation}
\begin{split}
&a_k^{(1 )} = \alpha_{k} a_k^{(2 )}- \beta_{k} a_{-k}^{(2 )\dagger} \\
&a_{-k}^{(1 )\dagger} =- \beta_{k} a_k^{(2 )}+ \alpha_{k} a_{-k}^{(2 )\dagger},
\end{split}  \label{BT1}
\end{equation}
and 
\begin{equation}
\begin{split}
&a_k^{(2 )} = \alpha_{k} a_k^{(1 )}+ \beta_{k} a_{-k}^{(1 )\dagger} \\
&a_{-k}^{(2 )\dagger} = \beta_{k} a_k^{(1)}+ \alpha_{k} a_{-k}^{(1 )\dagger},
\end{split}  \label{BT2}
\end{equation}
where
\begin{equation}
\begin{split}
\alpha_k=\frac{1}{2}\left(  \sqrt{ \frac{\omega_k^{(1)}}{\omega_k^{(2)}}} + \sqrt{\frac{\omega_k^{(2)}}{\omega_k^{(1)}}} \right), ~~~
\beta_k=-\frac{1}{2}\left(  \sqrt{ \frac{\omega_k^{(1)}}{\omega_k^{(2)}}} - \sqrt{\frac{\omega_k^{(2)}}{\omega_k^{(1)}}} \right). 
\end{split}  \label{wave function}
\end{equation}
From $a_k^{(2 )}\ket{0}_{2}=0$, we obtain
\begin{equation}
\begin{split}
a_k^{(1)} \ket{0}_2= \gamma_k a_{-k}^{(1)\dagger} \ket{0}_2 ,
\end{split}  \label{a to a-dagger}
\end{equation}
where
\begin{equation}
\begin{split}
\gamma_k \equiv - \frac{\beta_k}{\alpha_k}= \frac{\omega_k^{(1)}-\omega_k^{(2)}}{\omega_k^{(1)}+\omega_k^{(2)}} .
\end{split}  \label{wave functionb}
\end{equation}
We use the following notation, 
\begin{equation}
\begin{split}
\left< O \right>_{12} \equiv \frac{ _{1}\bra{0} O \ket{0}_2}{_{1}\braket{0|0}_{2}}, ~~~ 
\left< O \right>_{11} \equiv \frac{ _{1}\bra{0} O \ket{0}_1}{_{1}\braket{0|0}_{1}}, ~~~
\left< O \right>_{22} \equiv \frac{ _{2}\bra{0} O \ket{0}_2}{_{2}\braket{0|0}_{2}},
\end{split}  \label{wave functionc}
\end{equation}
where $O$ is an arbitrary operator. 
$\left< O \right>_{12}$ can be calculated as follows. 
First, we express $O$ as a function of $a_k^{(1)}$ and $a_k^{(1)\dagger}$ and represent it as the normal ordered operator. 
From ${}_1\bra{0} a_k^{(1)\dagger}=0$, $\left< O \right>_{12}$ can be expressed as a function of $\left< f(a_k^{(1)}) \right>_{12}$ where $f(a_k^{(1)})$ is a function of $a_k^{(1)}$. 

For later use, we consider
\begin{equation}
\begin{split}
F \equiv \sum_k f_{k} a_k^{(1)} ,
\end{split}  \label{wave functiond}
\end{equation}
where $f_{k}$ is an arbitrary complex function. 
By using (\ref{a to a-dagger}), we obtain 
\begin{equation}
\begin{split}
\left< F^2 \right>_{12} =  \sum_k \gamma_k f_k f_{-k},
\end{split}  \label{wave functiona}
\end{equation}
\begin{equation}
\begin{split}
\left< F^{2n+1} \right>_{12} = 0,
\end{split}  \label{matrix elements odd}
\end{equation}
and
\begin{equation}
\begin{split}
\left< F^{2n} \right> &= (2n-1) \left< F^2 \right>_{12}  \left< F^{2n-2} \right>_{12} \\
&= (2n-1)!! \left( \left< F^2 \right>_{12} \right)^{n}.
\end{split}  \label{matrix elements even}
\end{equation}
From (\ref{matrix elements odd}) and (\ref{matrix elements even}), we obtain
\begin{equation}
\begin{split}
\left< e^{F} \right>_{12} &= \sum_{n=0}^{\infty} \frac{1}{(2n)!} \left< F^{2n} \right>_{12} \\
&=\sum_{n=0}^{\infty} \frac{(2n-1)!!}{(2n)!}  \left( \left< F^2 \right>_{12} \right)^{n} \\
&= \sum_{n=0}^{\infty} \frac{1}{n!}  \left( \frac{1}{2} \left< F^2 \right>_{12} \right)^{n} \\
&= \exp \left[\frac{1}{2} \left< F^2 \right>_{12}  \right] . 
\end{split}  \label{e^F}
\end{equation}

\subsection{Operator method for Pseudo Entropy}\label{subsection OP}
We apply the operator method \cite{Shiba2014,Shiba2020} of entanglement entropy to the pseudo entropy. 
We review the operator method to compute the R\'{e}nyi entropy developed in \cite{Shiba2014}.
We consider $n$ copies of the scalar fields in $(d+1)$ dimensional spacetime and the $j$-th copy of the scalar field is denoted by 
$\{ \phi^{(j)} \}$. 
Thus the total Hilbert space, $H^{(n)}$, is the tensor product of the $n$ copies of the Hilbert space, 
$H^{(n)}= H \otimes H \dots \otimes H$ where $H$ is the Hilbert space of one scalar field. 
We define the density matrix $\rho^{(n)}$ in  $H^{(n)}$ as 
\begin{equation}
\rho^{(n)} \equiv \rho \otimes \rho \otimes \dots \otimes \rho
\end{equation}
where $\rho$ is an arbitrary density matrix in $H$. 
We can express $\mathrm{Tr} \rho_{\Omega}^n$ as 
\begin{equation}
\mathrm{Tr} \rho_{\Omega}^n=\mathrm{Tr} (\rho^{(n)} E_{\Omega}),  \label{expectation intro}
\end{equation}
where 
\begin{equation}
\begin{split}
E_{\Omega}=& \int \prod_{j=1}^{n} \prod_{a \in \Omega} \frac{dJ_{a}^{(j)}}{2\pi} dK_{a}^{(j)} \exp \left[i \sum_{l=1}^{n} \sum_{n\in \Omega} J_{n}^{(l+1)} \phi_{n}^{(l)} \right] \\
&\times  \exp \left[i \sum_{l=1}^{n} \sum_{n \in \Omega} K_{n}^{(l+1)} \pi_{n}^{(l)} \right] 
\exp \left[- i \sum_{l=1}^{n} \sum_{n \in \Omega} J_{n}^{(l)} \phi_{n}^{(l)} \right]
\end{split}  \label{glueing op}
\end{equation}
where $\pi_{n}^{(l)}$ is a conjugate momenta of $\phi_{n}^{(l)}$, $[\phi_{m}^{(l)}, \pi_{n}^{(l')}]=i\delta_{l, l'} \delta_{m, n}$, 
and $J^{(j)}_{n}$ and $K^{(j)}_{n}$ exist only in $\Omega$ and $J^{(n+1)}=J^{(1)}$.
Notice that $\phi$ and $\pi$ in (\ref{glueing op}) are operators and the ordering is important. 
This operator $E_{\Omega}$ is called as  \textit{the glueing operator}. 
When $\rho$ is a pure state, $\rho=\ket{\Psi}\bra{\Psi}$, 
the equation (\ref{expectation intro}) becomes 
\begin{equation}
\mathrm{Tr} \rho_{\Omega}^n = \bra{\Psi^{(n)}} E_{\Omega} \ket{\Psi^{(n)}}  \label{formula intro}
\end{equation}
where 
\begin{equation}
\ket{\Psi^{(n)}} =\ket{\Psi}\ket{\Psi}\dots \ket{\Psi} .
\end{equation}

The useful property of the glueing operator for calculating the pseudo entropy is the following property.
From eq.(2.18) in \cite{Shiba2014}, for $n$ arbitrary operators $F_{j} (j=1,2,\cdots n)$ on $H$, 
\begin{equation}
\begin{split}
\mathrm{Tr}(F_1 \otimes F_2 \otimes \cdots \otimes F_n \cdot E_{\Omega}) 
= \mathrm{Tr}(F_{1 \Omega}  F_{2\Omega}  \cdots  F_{n\Omega} ) 
\end{split}  \label{property}
\end{equation}
where $F_{j\Omega}\equiv \mathrm{Tr}_{\Omega^{c}} F_{j}$.

We consider the transition matrix,
\begin{equation}
\begin{split}
\tau^{1|2} \equiv \frac{\ket{0}_{2} {}_{1}\bra{0}}{{}_{1}\braket{0|0}_{2}}.
\end{split}  \label{wave functionn}
\end{equation}
By using the property (\ref{property}), we obtain 
\begin{equation}
\begin{split}
\mathrm{Tr} (\tau_{\Omega}^{1|2})^n = \frac{_{1}\bra{0^{(n)}} E_{\Omega} \ket{0^{(n)}}_2}{({}_{1}\braket{0|0}_{2})^n}
\end{split}  \label{trace 1}
\end{equation}
where $\ket{0^{(n)}}_{\alpha} = \ket{0}_{\alpha} \ket{0}_{\alpha} \cdots \ket{0}_{\alpha}, ~(\alpha=1,2)$.

In order to calculate $\mathrm{Tr} (\tau_{\Omega}^{1|2})^n$, we express $E_{\Omega}$ as a function of $a_k^{(1)}$ and $a_k^{(1)\dagger}$ and represent it as the normal ordered operator. 

We decompose $\phi$ and $\pi$ into the creation and annihilation parts,
\begin{equation}
\begin{split}
\phi_{n} = \phi_{n}^{1 +} + \phi_{n}^{1-},
\end{split}  \label{wave functionna}
\end{equation}
where
\begin{equation}
\begin{split}
& \phi_{n}^{1+} = \frac{1}{N^{d/2}} \sum_{k} \frac{1}{\sqrt{2\omega_k^{(1)}}} e^{2\pi i k n/N} a_{k}^{(1)} , ~~~
\phi_{n}^{1-} = (\phi_{n}^{1+})^{\dagger}, \\
&\pi_{n}^{1+} = \frac{1}{N^{d/2}} \sum_{k} (-i) \sqrt{\frac{\omega_k^{(1)}}{2}} e^{2\pi i k n/N} a_{k}^{(1)} , ~~~
\phi_{n}^{1-} = (\phi_{n}^{1+})^{\dagger}.
\end{split}  \label{wave functionnb}
\end{equation}
The commutators of these operators are
\begin{equation}
\begin{split}
&\left[\phi_{m}^{1+}, \phi_{n}^{1-} \right] = \left< \phi_{m} \phi_{n} \right>_{11} 
=  \frac{1}{N^d} \sum_k \frac{1 }{2 \omega_k^{(1)}} e^{2\pi i k (m-n)/N} 
\equiv X_{1, mn}, \\
&\left[\pi_{m}^{1+}, \pi_{n}^{1-} \right] = \left< \pi_{m} \pi_{n} \right>_{11} 
=  \frac{1}{N^d} \sum_k \frac{ \omega_k^{(1)} }{2} e^{2\pi i k (m-n)/N} 
\equiv P_{1, mn}, \\
&\left[\pi_{m}^{1+}, \phi_{n}^{1-} \right] =\left[\pi_{m}^{1-}, \phi_{n}^{1+} \right] 
=  \frac{1}{N^d} \sum_k \frac{ (-i) }{2} e^{2\pi i k (m-n)/N} 
=-\frac{i}{2} \delta_{m,n} \equiv Q_{mn}.
\end{split}  \label{commutators 1}
\end{equation}
By using (\ref{commutators 1}) and the Baker-Campbell-Hausdorff (BCH) formula $e^{X} e^{Y}=e^{[X,Y]} e^{Y} e^{X}, ~ e^{X+Y}=e^{-\frac{1}{2}[X,Y] } e^{X} e^{Y}$, for $[[X,Y],X]=[[X,Y],Y]=0$, we obtain
\begin{equation}
\begin{split}
E_{\Omega}=& \int \prod_{j=1}^{n} \prod_{a \in \Omega} \frac{dJ_{a}^{(j)}}{2\pi} dK_{a}^{(j)} \mathcal{N}_1 \left(\exp \left[i \sum_{l=1}^{n} \sum_{n \in \Omega} \left( \left( J_{n}^{(l+1)} - J_{n}^{(l)} \right) \phi_{n}^{(l)} + K_{n}^{(l)} \pi_{n}^{(l)} \right) \right] \right) \exp[-  S_{JK}]
\end{split}  \label{glueing op normal order}
\end{equation}
where $\mathcal{N}_1(O)$ is the normal ordered operator of $O$ with respect to $\phi_{n}^{1\pm}$ and $\pi_{n}^{1\pm}$, and
\begin{equation}
\begin{split}
S_{JK} = \sum_{l=1}^{n} \sum_{m, n \in \Omega} &\left[ \frac{1}{2} K_{m}^{(l)} P_{1,mn} K_{n}^{(l)} + \frac{1}{2} \left( J_{m}^{(l+1)} - J_{m}^{(l)} \right) X_{1,mn} \left( J_{n}^{(l+1)} - J_{n}^{(l)} \right) \right. \\ 
&\left. -K_{m}^{(l)} Q_{mn} \left( J_{n}^{(l)} + J_{n}^{(l+1)} \right) \right].
\end{split}  \label{S_JK}
\end{equation}

We substitute (\ref{glueing op normal order}) into (\ref{trace 1}) and obtain
\begin{equation}
\begin{split}
\mathrm{Tr} (\tau_{\Omega}^{1|2})^n &= \int \prod_{j=1}^{n} \prod_{a \in \Omega} \frac{dJ_{a}^{(j)}}{2\pi} dK_{a}^{(j)} 
\prod_{l=1}^{n}
\left< \exp \left[i  F^{(l)} \right] \right>_{12}
\exp[-  S_{JK}] \\
& =  \int \prod_{j=1}^{n} \prod_{a \in \Omega} \frac{dJ_{a}^{(j)}}{2\pi} dK_{a}^{(j)} 
\exp[-  S_{JK}- \frac{1}{2} \sum_{l=1}^{n} \left< (F^{(l)})^2 \right>_{12}],
\end{split}  \label{trace 2}
\end{equation}
where
\begin{equation}
\begin{split}
F^{(l)} \equiv \sum_{n \in \Omega} \left( \left( J_{n}^{(l+1)} - J_{n}^{(l)} \right) \phi_{n}^{1+} + K_{n}^{(l)} \pi_{n}^{1+} \right),
\end{split}  \label{wave functionnc}
\end{equation}
and we have used (\ref{e^F}). 
By using (\ref{a to a-dagger}), we obtain 
\begin{equation}
\begin{split}
& \frac{1}{2} \left<  (F^{(l)})^2  \right>_{12} \\
& = 
\sum_{l=1}^{n} \sum_{m, n \in \Omega} \left[ \frac{1}{2} K_{m}^{(l)} \bar{P}_{mn} K_{n}^{(l)} + \frac{1}{2} \left( J_{m}^{(l+1)} - J_{m}^{(l)} \right) \bar{X}_{mn} \left( J_{n}^{(l+1)} - J_{n}^{(l)} \right) +K_{m}^{(l)} R_{mn} \left( J_{n}^{(l+1)} - J_{n}^{(l)} \right) \right] ,
\end{split}  \label{F square}
\end{equation}
where 
\begin{equation}
\begin{split}
\bar{X}_{mn}\equiv \left< \phi_{m}^{1+} \phi_{n}^{1+} \right>_{12} = \frac{1}{N^d} \sum_k \frac{\gamma_k }{2 \omega_k^{(1)}} e^{2\pi i k (n-m)/N} ,
\end{split}  \label{wave functionw}
\end{equation}
\begin{equation}
\begin{split}
\bar{P}_{mn}\equiv \left< \pi_{m}^{1+} \pi_{n}^{1+} \right>_{12} = \frac{1}{N^d} \sum_k \frac{-\omega_k^{(1)}\gamma_k }{2 } e^{2\pi i k (n-m)/N} ,
\end{split}  \label{wave functionp}
\end{equation}
\begin{equation}
\begin{split}
R_{mn} &\equiv \left< \phi_{m}^{1+} \pi_{n}^{1+} \right>_{12} =\frac{1}{2} \left< (\phi_{m}\pi_{n}+ \pi_{m}\phi_{n}) \right>_{12} \\
&= \frac{1}{N^d} \sum_k \frac{-i\gamma_k }{2 } e^{2\pi i k (m-n)/N} 
= \frac{1}{N^d} \sum_k \frac{(-i)  }{2 } \frac{\omega_k^{(1)}-\omega_k^{(2)}}{\omega_k^{(1)}+\omega_k^{(2)}} e^{2\pi i k (m-n)/N} .
\end{split}  \label{wave functionq}
\end{equation}
From (\ref{S_JK}), (\ref{trace 2}) and (\ref{F square}), we obtain
\begin{equation}
\begin{split}
&S_{12,JK} \equiv S_{JK}+ \frac{1}{2} \sum_{l=1}^{n}  \left< (F^{(l)})^2 \right>_{12} \\
&= \sum_{l=1}^{n} \sum_{m, n \in \Omega} \left[ \frac{1}{2} K_{m}^{(l)} P_{mn} K_{n}^{(l)} + \frac{1}{2} \left( J_{m}^{(l+1)} - J_{m}^{(l)} \right) X_{mn} \left( J_{n}^{(l+1)} - J_{n}^{(l)} \right) \right. \\
&\left. K_{m}^{(l)} \left(-Q_{mn} \left( J_{n}^{(l)} + J_{n}^{(l+1)}  \right) +R_{mn} \left( J_{n}^{(l+1)} - J_{n}^{(l)}  \right) \right) \right], 
\end{split}  \label{S_12JK}
\end{equation}
where 
\begin{equation}
\begin{split}
X_{mn}&\equiv X_{1,mn}+ \bar{X}_{mn}= \left< \phi_{m} \phi_{n} \right>_{12} \\
&= \frac{1}{N^d} \sum_k \frac{(1+\gamma_k) }{2 \omega_k^{(1)}} e^{2\pi i k (m-n)/N} 
= \frac{1}{N^d} \sum_k \frac{1 }{ \omega_k^{(1)}+\omega_k^{(2)}} e^{2\pi i k (m-n)/N},
\end{split}  \label{wave functionr}
\end{equation}
\begin{equation}
\begin{split}
P_{mn}&\equiv P_{1,mn}+ \bar{P}_{mn}= \left< \pi_{m} \pi_{n} \right>_{12} \\
&= \frac{1}{N^d} \sum_k \frac{\omega_k^{(1)} (1-\gamma_k)}{2 } e^{2\pi i k (m-n)/N} 
= \frac{1}{N^d} \sum_k \frac{\omega_k^{(1)} \omega_k^{(2)} }{ \omega_k^{(1)}+ \omega_k^{(2)}} e^{2\pi i k (m-n)/N} .
\end{split}  \label{wave functions}
\end{equation}


We perform the $J$ and $K$ integrals in (\ref{trace 2}) simultaneously.
We rewrite $S_{12,JK}$ in (\ref{S_12JK}) as
\begin{equation}
\begin{split}
S_{12,JK}=S_{12,J} = (J^{(1)T},\cdots, J^{(n)T}, K^{(1)T},\cdots, K^{(n)T}) \bar{S}_n
\begin{pmatrix}
J^{(1)}  \\
\vdots \\
J^{(n)} \\
K^{(1)} \\
\vdots \\
K^{(n)}
\end{pmatrix},
\end{split}  \label{S_12,JK2}
\end{equation}
where
\begin{equation}
\bar{S}_n =
\begin{pmatrix}
S_{nJJ} & S_{nJK}   \\
S_{nKJ}& S_{nKK}  \\
\end{pmatrix} , \label{S_n}
\end{equation}
and,
\begin{equation}
\begin{split}
&(S_{nJJ})_{l,l'} = X \delta_{l,l'} -\frac{X}{2} \delta_{l,l'-1} -\frac{X}{2} \delta_{l,l'+1}, \\
&(S_{nKK})_{l,l'} = \frac{P}{2} \delta_{l,l'} , \\
&(S_{nJK})_{l,l'} = -\frac{1}{2}(Q+R) \delta_{l,l'}-\frac{1}{2}(Q-R) \delta_{l,l'+1}, \\
&(S_{nKJ})_{l,l'} = -\frac{1}{2}(Q+R) \delta_{l,l'}-\frac{1}{2}(Q-R) \delta_{l,l'-1},
\end{split}
\label{barS_n elements}
\end{equation}
where $\delta_{1,n+1}=\delta_{n,0}=1$.
We substitute (\ref{S_12,JK2}) into (\ref{trace 2}) and perform the $J$ and $K$ integrals in (\ref{trace 2}) and obtain
\begin{equation}
\begin{split}
\mathrm{Tr} (\tau_{\Omega}^{1|2})^n 
& = \left(\det 2 \bar{S}_{n}\right)^{-1/2}  .
\end{split}  \label{trace 6}
\end{equation}

We can diagonalize $S_n$ with respect to the replica label $l$ by Fourier transformation. 
We define a unitary matrix $U_{lk}=\frac{1}{\sqrt{n}}e^{i2\pi kl/n}$ and obtain, 
\begin{equation}
\begin{pmatrix}
U^{\dagger} & 0  \\
0& U^{\dagger}  \\
\end{pmatrix}
\begin{pmatrix}
S_{nJJ} & S_{nJK}   \\
S_{nKJ}& S_{nKK}  \\
\end{pmatrix} 
\begin{pmatrix}
U & 0  \\
0& U \\
\end{pmatrix}
= \begin{pmatrix}
U^{\dagger} S_{nJJ} U & U^{\dagger} S_{nJK} U  \\
U^{\dagger} S_{nKJ} U & U^{\dagger} S_{nKK} U \\
\end{pmatrix}    , \label{barS_n fourier}
\end{equation}
where 
\begin{equation}
\begin{split}
&(U^{\dagger} S_{nJJ} U)_{k,k'} = X (1-\cos \frac{2\pi k}{n})\delta_{k,k'}  \\
&(U^{\dagger} S_{nKK} U)_{k,k'} = \frac{P}{2} \delta_{k,k'} , \\
&(U^{\dagger} S_{nJK} U )_{k,k'} = -\frac{1}{2}((Q+R)+(Q-R)e^{-i2\pi k/n}) \delta_{k,k'}, \\
&(U^{\dagger} S_{nKJ} U)_{k,k'} = -\frac{1}{2}((Q+R)+(Q-R)e^{i2\pi k/n}) \delta_{k,k'}.
\end{split}
\label{barS_n elements fourier}
\end{equation}
From (\ref{barS_n fourier}) and (\ref{barS_n elements fourier}), we obtain 
\begin{equation}
\begin{split}
\det 2\bar{S}_{n} = \prod_{k=0}^{n-1} \det 2S_{n,k}  ,
\end{split}
\label{det S}
\end{equation}
where
\begin{equation}
 S_{n,k}=
\begin{pmatrix}
X (1-\cos \frac{2\pi k}{n}) & -\frac{1}{2}((Q+R)+(Q-R)e^{-i2\pi k/n})  \\
-\frac{1}{2}((Q+R)+(Q-R)e^{i2\pi k/n}) & \frac{P}{2}  \\
\end{pmatrix}   . \label{S_n,k}
\end{equation}
For $k=0$, we obtain
\begin{equation}
\det 2S_{n,k=0} = \det P \det P^{-1} =1    . \label{S_n,ka}
\end{equation}
where we used the formula
\begin{equation}
\det
\begin{pmatrix}
A & B \\
C & D \\
\end{pmatrix}  
=\det(A-BD^{-1}C) \det D . \label{schur complement}
\end{equation}
For $k \geq 1 $, we can rewrite $S_{n,k}$ as
\begin{equation}
2S_{n,k}=
\begin{pmatrix}
- (1-e^{-i2\pi k/n}) & 0  \\
0 & 1 \\
\end{pmatrix} 
2\tilde{S}_{n,k}
\begin{pmatrix}
- (1-e^{i2\pi k/n}) & 0  \\
0 & 1 \\
\end{pmatrix}  . \label{S_n,k 2}
\end{equation}
where
\begin{equation}
2\tilde{S}_{n,k} = \Gamma -\frac{1}{2} \cot \frac{\pi k}{n} J 
=iJ \left( iJ \Gamma + \frac{i}{2} \cot \frac{\pi k}{n} \right)  , \label{tildeS_n,k}
\end{equation}
here
\begin{equation}
\Gamma =
\begin{pmatrix}
X & R  \\
R & P \\
\end{pmatrix},  ~~~
J=
\begin{pmatrix}
0 & 1  \\
-1 & 0 \\
\end{pmatrix}  , \label{S_n,kc}
\end{equation}
and we used $Q=-\frac{i}{2}$. 
In order to calculate $\prod_{k=1}^{n-1} \det 2S_{n,k}$, we use the following formulas, (we show them in the next subsection),
\begin{equation}
\begin{split}
&\prod_{k=1}^{n-1} (1-ye^{i2\pi k/n}) =\frac{1-y^n}{1-y} , \\
&\prod_{k=1}^{n-1} (1-e^{i2\pi k/n}) =n , \\
&\prod_{k=1}^{n-1} \left(z+ \frac{i}{2} \cot \frac{\pi k}{n} \right) 
= \frac{1}{n} \left[ \left(z+\frac{1}{2}\right)^n - \left(z-\frac{1}{2}\right)^n \right]  . 
\end{split}
\label{formulas}
\end{equation}
From (\ref{S_n,k 2}), (\ref{tildeS_n,k}) and (\ref{formulas}), we obtain
\begin{equation}
\begin{split}
\det 2\bar{S}_{n} &= \left(\det iJ \right)^{n-1} \det \left[\left(iJ\Gamma+\frac{1}{2}\right)^n -\left(iJ\Gamma-\frac{1}{2}\right)^n \right] \\
&= (-1)^{V(n-1)} \prod_{i=1}^{2V} \left[\left(\nu_i+\frac{1}{2}\right)^n -\left(\nu_i-\frac{1}{2}\right)^n \right]  \\
&= \prod_{i=1}^{V} \left[\left(\nu_i+\frac{1}{2}\right)^n -\left(\nu_i-\frac{1}{2}\right)^n \right] 
\prod_{i=V+1}^{2V} \left[\left(-\nu_i+\frac{1}{2}\right)^n -\left(-\nu_i-\frac{1}{2}\right)^n \right]  ,
\end{split}
\label{det S_n}
\end{equation}
where $\nu_i$ is the eigenvalue of $iJ\Gamma$ and $V$ is the number of the points of the subsystem.
From the characteristic equation, we obtain $0=\det (x-iJ\Gamma) = \det (x-iJiJ\Gamma iJ)=\det (x-\Gamma iJ)=\det (x+(iJ\Gamma)^{T})=\det (x+iJ\Gamma)$, where $x$ is an eigenvalue of $iJ\Gamma$ and we used $(iJ)^2=1$ and $\Gamma=\Gamma^{T}$. 
So, if $x$ is an eigenvalue of $iJ\Gamma$,  $-x$ is also an an eigenvalue of $iJ\Gamma$. 
So, we sort $\nu_i$ as $\nu_{V+i}=-\nu_{i}$ and obtain
\begin{equation}
\begin{split}
\det 2\bar{S}_{n} = \prod_{i=1}^{V} \left[\left(\nu_i+\frac{1}{2}\right)^n -\left(\nu_i-\frac{1}{2}\right)^n \right]^{2} 
\end{split}
\label{det S_n 2}
\end{equation}

From (\ref{trace 6}) and (\ref{det S_n 2}), we obtain the pseudo (R\'{e}nyi) entropy as,
\begin{equation}
\begin{split}
&S^{(n)}(\tau_\Omega^{1|2}) =\frac{1}{1-n} \ln \mathrm{Tr}\left[ (\tau_\Omega^{1|2})^n \right] \\
&=\frac{1}{n-1} 
 \sum_{i=1}^{V} \ln \left[\left(\nu_i+\frac{1}{2}\right)^n -\left(\nu_i-\frac{1}{2}\right)^n \right] ,
\end{split}  \label{Renyi PE 1}
\end{equation}
\begin{equation}
\begin{split}
&S(\tau_\Omega^{1|2}) = \lim_{n\to 1} S^{(n)}(\tau_\Omega^{1|2}) \\
&= 
 \sum_{i=1}^{V} \left[ \left(\nu_i+\frac{1}{2}\right) \ln \left(\nu_i+\frac{1}{2}\right) - \left(\nu_i-\frac{1}{2}\right) \ln \left(\nu_i-\frac{1}{2}\right) \right] . \\
\end{split}  \label{vN PE 1}
\end{equation}
When $R=0$, $\Gamma$ is a positive-definite real matrix and we can show that $-\nu_{V+i}=\nu_i \geq 0$ by using Williamson's theorem. 
So, when $R=0$, eqs (\ref{Renyi PE 1}) and (\ref{vN PE 1}) are the same as ordinary entanglement (R\'{e}nyi) entropy.

\subsection{proof of (\ref{formulas})} \label{proof}
We use the following formula (p.25 in \cite{Iwanami2}),
\begin{equation}
\begin{split}
&\prod_{r=1}^{n}(1+a^r x) = 1+ \sum_{r=1}^{n} \frac{(1-a^n)(1-a^{n-1}) \cdots (1-a^{n-r+1})}{(1-a)(1-a^2)\cdots (1-a^r)} a^{r(r+1)/2}x^r .
\end{split}  \label{formula 1}
\end{equation}
From (\ref{formula 1}), we obtain
\begin{equation}
\begin{split}
&\prod_{k=1}^{n-1} (1-ye^{i2\pi k/n}) =\frac{1-y^n}{1-y}  . 
\end{split}
\label{formulas 1}
\end{equation}
By taking the $y\to1$ limit, we obtain 
\begin{equation}
\begin{split}
&\prod_{k=1}^{n-1} (1-e^{i2\pi k/n}) = n . 
\end{split}
\label{formulas 2}
\end{equation}
We can rewrite $z+ \frac{i}{2} \cot \frac{\pi k}{n}$ as 
\begin{equation}
\begin{split}
&z+ \frac{i}{2} \cot \frac{\pi k}{n} = \frac{z+1/2}{(1-e^{i2\pi k/n})} 
\left(1-\frac{z-1/2}{z+1/2} e^{i2\pi k/n} \right) .
\end{split}
\label{formulas 3 pre}
\end{equation}
From (\ref{formulas 1}), (\ref{formulas 2}) and (\ref{formulas 3 pre}), we obtain
\begin{equation}
\begin{split}
&\prod_{k=1}^{n-1} \left(z+ \frac{i}{2} \cot \frac{\pi k}{n} \right) 
= \frac{1}{n} \left[ \left(z+\frac{1}{2}\right)^n - \left(z-\frac{1}{2}\right)^n \right]  . 
\end{split}
\label{formulas 3}
\end{equation}

\section{Appendix C: Almost Massless Regimes}


In this section, we consider a periodic system with length $L$ and `almost massless' scalar fields with mass $m_i L\ll1$. Let $\rho^{i}_{A}$ be a reduced density matrix for an almost massless scalar field in a single interval $A=[0, l]$. It is known that the entanglement entropy for $\rho^{i}_{A}$ is schematically given by
\be
S(\rho^{i}_{A})=\dfrac{c}{3}\log\left[\dfrac{L}{\pi \epsilon}\sin\left(\frac{\pi l}{L}\right)\right]-\dfrac{1}{2}\log (m_iL)+f(m_i,L, l), \label{eq:nthRenyi}
\ee
where $f(m_i,L, l)$ is a non-trivial function which is negligible in our almost massless field and is not important for our present discussion. 

\begin{figure}[t]
 \begin{center}
 \resizebox{60mm}{!}{
 \includegraphics{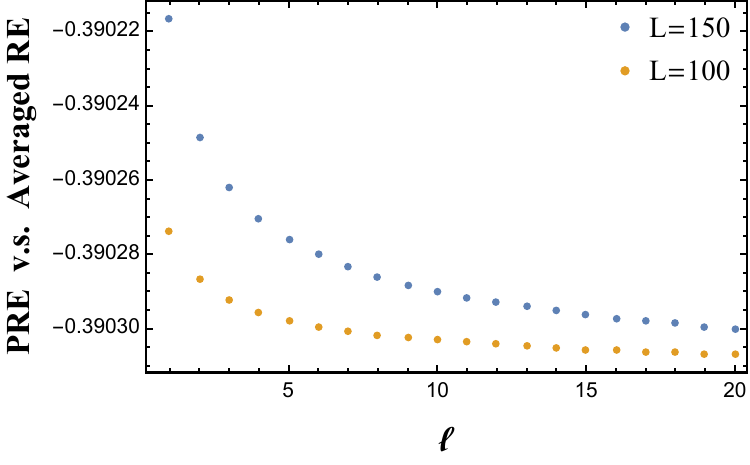}}
 \end{center}
 \caption{Difference between the 2nd pseudo Renyi entropy and the averaged value of 2nd the Renyi entropy. Here we set $m_1=1.0\times 10^{-5}$ and $m_2=1.7\times 10^{-4}$. Note that the pseudo Renyi entropy is smaller than the averaged value of ordinary ones. We have small $l$-dependence but it is negligible up to $3$ or $4$ digit. It means that the second term of \eqref{eq:PEscalar} essentially explains this negative value.}\label{fig:diff}
\end{figure}

On the other hand, for the pseudo entropy for two almost massless scalar fields with mass $m_i$ and $m_j$, we numerically confirmed
\be
S(\tau^{i|j}_A)=\dfrac{c}{6}\log\left[\dfrac{L}{\pi \epsilon}\sin\left(\frac{\pi l}{L}\right)\right]-\dfrac{1}{2}\log \left[\frac{m_i+m_j}{2} L\right]+f_0(m_i,m_j,L, l), \label{eq:PEscalar}
\ee
where $f_0(m_i,m_j,L, l)$ is again a negligible function which is less important than the second term in \eqref{eq:PEscalar}. In particular, we have numerically studied the difference between the pseudo entropy and the averaged entanglement entropy,
\be
\Delta S_{12}\equiv S(\tau^{1|2}_A)-\frac{S(\rho^{1}_{A})+S(\rho^{2}_{A})}{2}.
\ee
Interestingly, it can be well-approximated as the mass terms in the above,
\be
\Delta S_{12}\simeq -\dfrac{1}{2}\log \left[\frac{m_1+m_2}{2} L\right]+\dfrac{1}{2}\left(\dfrac{1}{2}\log (m_1L)+\dfrac{1}{2}\log (m_2L)\right)=-\dfrac{1}{4}\log\left[\dfrac{(m_1+m_2)^2}{4m_1m_2}\right],
\ee
which does not depend on the system size. Notice that it is always negative in our almost massless regimes. It means that these mass terms (the second term of \eqref{eq:nthRenyi} and \eqref{eq:PEscalar}) essentially explain the negativity of $\Delta S_{12}$. For massive regions, however, we cannot neglect the third terms of these equations and still observe the negativity of $\Delta S_{12}$. We have confirmed the same behaviour for the 2nd pseudo Renyi entropy. See FIG. \ref{fig:diff}. 

\subsection*{Lifshitz cases with $z_1=z_2>1$}
\begin{figure}[t]
 \begin{center}
 \resizebox{60mm}{!}{
 \includegraphics{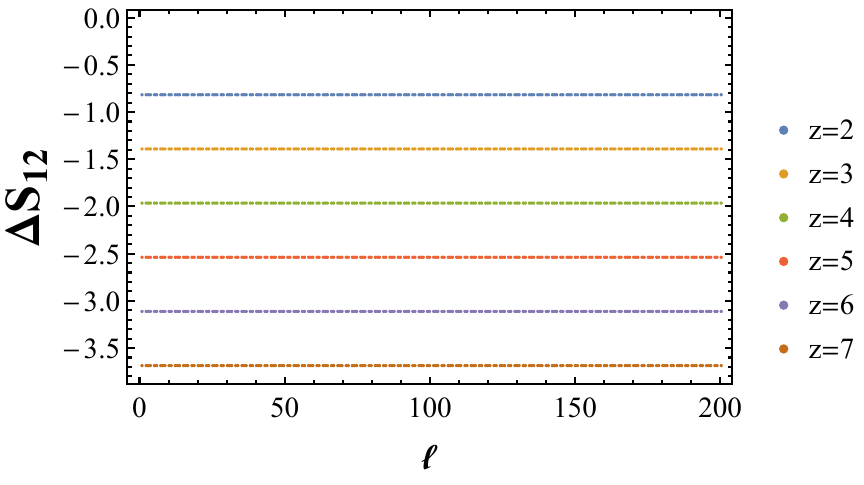}}
 \end{center}
 \caption{The $z$-dependence of the difference between pseudo entropy and averaged entanglement entropies, $\Delta S_{12}$. Here we set $L=2000, m_1=1.0\times 10^{-7}, m_2=1.0\times 10^{-8}$ and $z_1=z_2\equiv z$. We stress that these are not evenly spaced and it can be perfectly explained by the equation \eqref{eq:liflifzz}. We have seen this agreement up to $16$ digits. Notice that we did not see such an almost perfect coincidence for $z=1$ case.}\label{fig:barediff}
\end{figure}

One can repeat the same analysis for $z_1=z_2\equiv z>1$ cases and ask a $z$-dependence of the previous mass-terms. 
The answer is simply given by replacing $m_{1,2}L$ to $(m_{1,2}L)^z$. 
To be explicit, we have numerically confirmed
\be
\Delta S_{12}\simeq -\dfrac{1}{2}\log \left[\frac{(m_1L)^z+(m_2L)^z}{2}\right]+\dfrac{1}{2}\left(\dfrac{z}{2}\log (m_1L)+\dfrac{z}{2}\log (m_2L)\right)=-\dfrac{1}{4}\log\left[\dfrac{(m_1^z+m_2^z)^2}{4(m_1m_2)^z}\right]. \label{eq:liflifzz}
\ee
We stress that the $z$-dependence of the pseudo entropy does not show up as an overall factor (see FIG. \ref{fig:barediff}).

\section{Appendix D: Massive Regimes}


In this appendix, we study the pseudo entropy for massive scalar fields. In contrast to the previous almost massless regime explained in appendix C, our result is based on semi-analytic approach. We will leave the detail of the calculation in the end of this appendix.
Based on our correlator method, we propose a mass-correction formula of the pseudo entropy for scalar fields as
\begin{align}
S(\tau^{1|2}_{A_l})-S(\tau^{1|2}_{A_{l_0}})&=f(m_1,m_2, l)-f(m_1,m_2, l_0),\label{eq:regPE}
\end{align}
where
\begin{align}
f(m_1,m_2, l)&=\dfrac{1}{3}\log\left[\dfrac{L}{\epsilon \pi}\sin\left(\dfrac{\pi  l}{L}\right)\right]+\dfrac{1}{2}\log\left[-\dfrac{m_1^2\log[m_1 l]-m_2^2\log[m_2 l]}{m_1^2-m_2^2}\right]. 
\end{align}
Here $S(\tau^{1|2}_{A_l})$ gives the pseudo entropy for a single interval $A_l=[0,l]$ between two vacua with different mass parameters $m_1$ and $m_2$. The $l_0$ is just a reference point to get rid of irrelevant contributions. Note that this formula is a leading order approximation and only valid for the small interval, $m_1l, m_2l\ll1$. Under the appropriate limit with $m_1\rightarrow m_2$, it reduces to the famous result for the entanglement entropy for a massive scalar field\cite{Casini:2005zv}.

Notice that the $f(m_1,m_2,l)$ is symmetric, {\it i.e.} $f(m_1,m_2,l)=f(m_2,m_1,l)$ which is also guaranteed by our numerical results. On the other hand, we have to mention that the $l$-dependence of $f(m_1,m_2,l)$ is not sensitive to the mass parameters very much. 

\begin{figure}[h]
 \begin{center}
  \resizebox{120mm}{!}{
 \includegraphics{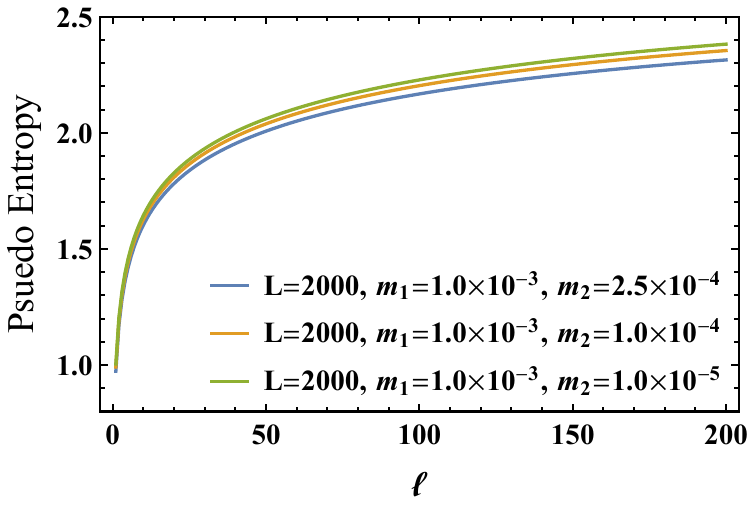}
 \includegraphics{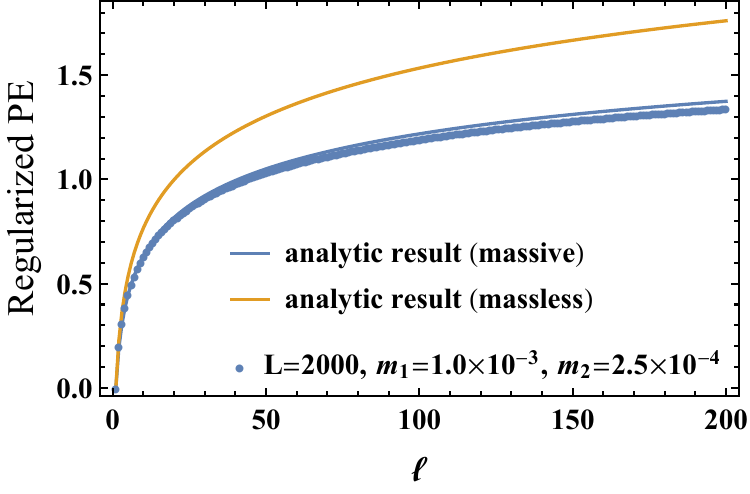}}
 \resizebox{120mm}{!}{
 \includegraphics{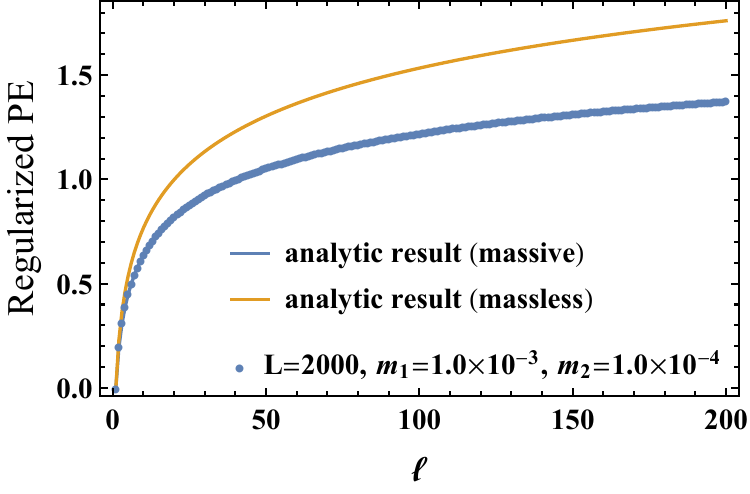}
 \includegraphics{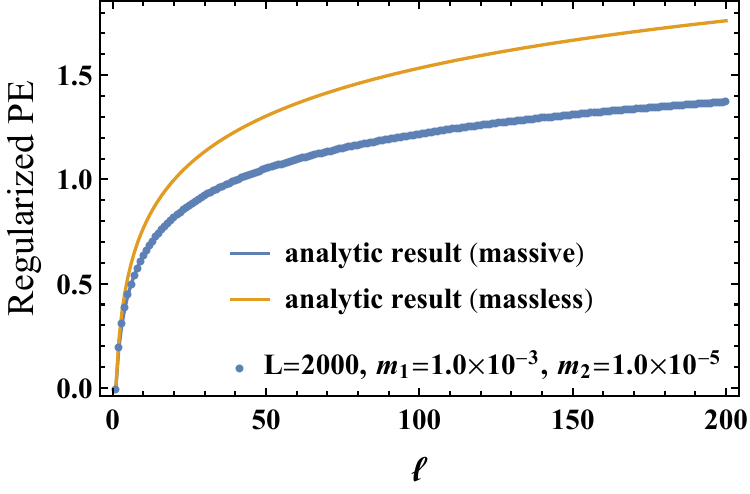}}
 \end{center}
 \caption{The PE and regularized PE for fixed $m_1=1.0\times 10^{-3}$ with various mass parameters $m_2$. As a reference, we also plot the entanglement entropy for CFT vacuum with $c=1$ (orange curve). Note that this formula is valid only in the regime $m_il\ll1$. Out of this regime, as we can see from the right-top figure, there is a small deviation. }
 \label{fig:dep_mass}
\end{figure}

For convenience, we define a  regularized PE as
\begin{equation}
    S_{\textrm{reg.}}(\tau^{1|2}_{A_l})=S(\tau^{1|2}_{A_l})-S(\tau^{1|2}_{A_{\epsilon}}),
\end{equation}
which corresponds to the left hand side of \eqref{eq:regPE} with $l_0=\epsilon$. In Figure \ref{fig:dep_mass}, we plotted PE and regularized PE for fixed $m_1$ with various mass parameters $m_2$. These figures numerically guarantee that the above mass-corrected formula is valid. 

\begin{figure}[h]
 \begin{center}
  \resizebox{70mm}{!}{
 \includegraphics{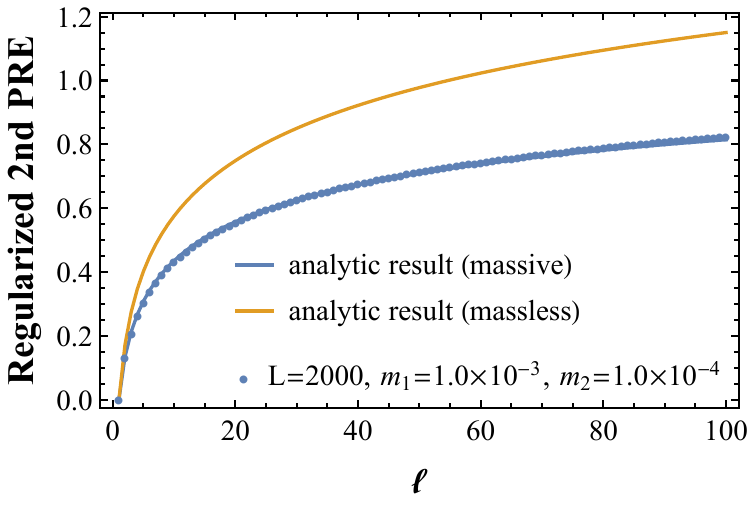}}
 \end{center}
 \caption{The same plot as FIG. \ref{fig:dep_mass} for the regularized 2nd PRE.}
 \label{fig:dep_mass2}
\end{figure}

In the same way, we can also find the similar expression for 2nd pseudo Renyi entropy as, 
\begin{align}
S^{(2)}(\tau^{1|2}_{A_l})-S^{(2)}(\tau^{1|2}_{A_{l_0}})&=g(m_1,m_2,l)-g(m_1,m_2,l_0),\label{eq:regPEa}
\end{align}
where
\begin{align}
g(m_1,m_2,l)&=\dfrac{1}{4}\log\left[\dfrac{L}{\epsilon \pi}\sin\left(\dfrac{\pi l}{L}\right)\right]+\dfrac{1}{2}\log\left[-\dfrac{m_1^2\log[m_1l]-m_2^2\log[m_2l]}{m_1^2-m_2^2}\right]. 
\end{align}
Note that the mass correction part does not depend on the Renyi index as well as the ordinary entanglement entropy. To see the consistency with numerical results, please see the Figure \ref{fig:dep_mass2}.

\subsection{Detail of the semi-analytic derivation}\label{app:massive}
In what follows, we explain a semi-analytic derivation of the above mentioned mass-correction formula from our covariance matrix methods. 

A key idea is to notice that the mass-dependence is an IR effect which can be read off from the low energy modes in the discretized models.
Having this intuition, let us treat a single site on the lattice as our subsystem and only focus on the lowest energy mode in the dispersion relation. The similar approach has been accomplished in \cite{Chapman:2018hou, MozaffarMollabashi:2020}.
That is to say, we take the thermodynamic limit $N\rightarrow\infty$ and approximate our dispersion relation as,
\be
\omega^{(i)}_k=\sqrt{m^2_i+\frac{4}{\epsilon^2}\sin^2\frac{\pi k}{N}}\simeq \sqrt{m^2_i+\frac{4}{\epsilon^2}\left(\frac{\pi k}{N}\right)^2},
\ee
where we recovered the lattice size $\epsilon$ which now formally coincides with the subsystem size $l$. 
In this limit, each component of the matrix becomes an integral form,
\begin{align}
X_{11}&\simeq\dfrac{\epsilon}{4\pi}\int^{\frac{2\pi}{\epsilon}}_{0}dp\,\dfrac{2}{\omega^{(1)}_p+\omega^{(2)}_p},\\
P_{11}&\simeq\dfrac{\epsilon}{4\pi}\int^{\frac{2\pi}{\epsilon}}_{0}dp\, \dfrac{2\omega^{(1)}_p\omega^{(2)}_p }{\omega^{(1)}_p+\omega^{(2)}_p},\\
R_{11}&\simeq \dfrac{i\epsilon}{4\pi}\int^{\frac{2\pi}{\epsilon}}_{0}dp\, \dfrac{\omega^{(1)}_p-\omega^{(2)}_p }{\omega^{(1)}_p+\omega^{(2)}_p},
\end{align}
where each $\omega^{(i)}_p$ follows the standard dispersion relation of a massive free scalar field,
\be
\omega^{(i)}_p=\sqrt{m^2_i+p^2}.
\ee
Following our prescription, we shall study the eigenvalue of our covariance matrix,
\be
\bpm
X_{11} & R_{11} \\
R_{11} & P_{11}
\epm.
\ee
We can formally expand each component with respect to the small $\epsilon$. Physically, we have to assume $m_{1,2}\epsilon\ll 1$. Remind that now we can regard $\epsilon$ as a subsystem size $l$. In doing so, we obtain the leading contribution of interest,
\begin{align}
X_{11}&=\dfrac{\epsilon}{8\pi}+\dfrac{\epsilon}{8(m_1^2-m_2^2)\pi}\left(m_1^2\log\left(\frac{16\pi^2}{m_1^2\epsilon^2}\right)-m_2^2\log\left(\frac{16\pi^2}{m_2^2\epsilon^2}\right)\right)+\mathcal{O}(\epsilon),\\
P_{11}&=\dfrac{\pi}{2\epsilon}+\mathcal{O}(\epsilon),\\
R_{11}&=\mathcal{O}(\epsilon).
\end{align}
In particular, we can neglect the off-diagonal elements $R_{11}$ up to this order.
It means that we can simply obtain the desired eigenvalue $\nu$ as
\begin{align}
\nu&\simeq\sqrt{X_{11}P_{11}}\simeq\sqrt{\dfrac{1}{16}+\dfrac{1}{8}\dfrac{m_1^2\log[\frac{4\pi}{m_1\epsilon}]-m_2^2\log[\frac{4\pi}{m_2\epsilon}]}{m_1^2-m_2^2}}. 
\end{align}
Finally, we have obtained the analytic expression of the pseudo entropy as
\begin{align}
S(\tau^{1|2}_{A_\epsilon}) \simeq \log(\nu)\simeq\dfrac{1}{2}\log\left(-\dfrac{m_1^2\log[m_1\epsilon]-m_2^2\log[m_2\epsilon]}{m_1^2-m_2^2}\right). \label{eq:massive_app}
\end{align}
As a consistency check, it is symmetric under the mass exchange $m_1\leftrightarrow m_2$ and reduces to the well-known formula by Casini and Heurta under the ordinary entropy limit $m_1\rightarrow m_2$. As we have already seen in FIG. \ref{fig:dep_mass}, this expression matches the numerical calculations. 

In the similar way, one can also consider the similar analytic form for any $n$-th Renyi entropy. For example, if we consider the 2nd pseudo Renyi entropy, we obtain the same form as \eqref{eq:massive_app},
\be
S^{(2)}(\tau^{1|2}_{A_\epsilon})=\log2\nu\simeq \dfrac{1}{2}\log\left(-\dfrac{m_1^2\log[m_1\epsilon]-m_2^2\log[m_2\epsilon]}{m_1^2-m_2^2}\right). 
\ee
which has the same form as \eqref{eq:massive_app} if we focus on the leading order contribution and is consistent with the numerical plots (see FIG.\ref{fig:dep_mass2}).  

Our approach nicely captures the leading order of mass-corrections. Finding more refined or exact analytical approaches would be an interesting future direction.

\section{Appendix E: Pseudo Entropy under Small Perturbations}
Here we work out the behavior of the pseudo entropy $S(\tau^{1|2}_A)$ when 
the reduced transition matrix $\tau^{1|2}_A$ is changed infinitesimally.

\subsection{First Law of Pseudo Entropy}
Consider two transition matrices $\tau$ and $\tau_0$, which are very closed to each other.
We write the difference as $\tau-\tau_0=\delta \tau$.
We consider a generalization of relative entropy to the transition matrices defined by
\ba
S(\tau|\tau_0)=\mbox{Tr}[\tau\log\tau]-\mbox{Tr}[\tau\log\tau_0].
\ea
We expand $S(\tau|\tau_0)$ up to the quadratic order as follows (note the relation $S(\tau_0|\tau_0)=0$):
\ba
S(\tau|\tau_0)\simeq \int^\infty_0 dt\ \mbox{Tr}\left[\frac{t}{(t+\tau_0)^2}\delta\tau
\frac{1}{t+\tau_0}\delta\tau\right].  \label{quadp}
\ea
Note that if $\tau_0$ is non-negative as in the ordinary density matrices, the above quadratic term is positive. 

Let us rewrite $S(\tau|\tau_0)$ as follows:
\ba
&&S(\tau|\tau_0)=S(\tau_0)-S(\tau)+\mbox{Tr}[\tau H]-\mbox{Tr}[\tau_0 H],
\ea
where $H=-\log \tau_0$ is a `pseudo' modular Hamiltonian. 
Since in the linear order approximation $O(\delta \tau)$, the relative pseudo 
entropy is vanishing, we obtain the first law:
\ba
S(\tau)-S(\tau_0)\simeq \la H\lb_{\tau}-\la H\lb_{\tau_0}+O(\delta\tau^2).  \label{fstlp}
\ea
This can be regarded as a generalization of the first law of entanglement entropy
 \cite{Bhattacharya:2012mi,Blanco:2013joa,Wong:2013gua}.

If we include the quadratic order (\ref{quadp}), we have 
\ba
&&S(\tau)-S(\tau_0)\simeq \la H\lb_{\tau}-\la H\lb_{\tau_0}
-\int^\infty_0 dt\ \mbox{Tr}\left[\frac{t}{(t+\tau_0)^2}\delta\tau
\frac{1}{t+\tau_0}\delta\tau\right].  \label{pertgh}
\ea
The final integral term is negative if $\tau_0$ is non-negative and $\delta\tau$ is hermitian.

Consider two quantum states $|\psi_1\lb$ and $|\psi_2\lb$ which are both very close to a 
state $|\psi_0\lb$. 
In this case the deviation of the pseudo entropy $\tau^{\psi_1|\psi_2}_A$ from $S(\rho^0_A)$ is found from the first law (\ref{fstlp}):
\ba
S(\tau^{1|2}_A)-S(\rho^0_A)\simeq \frac{\la \psi_2|H_A|\psi_1\lb}{\la \psi_2|\psi_1\lb}+O(\delta \tau^2). \label{perty}
\ea
Here $H_A$ is the modular Hamiltonian defined by $H_A=-\log\rho_A+S(\rho^0_A)$ such that 
$\la \psi_0|H_A|\psi_0\lb=0$.  For example, we can regard $|\psi_0\lb$ as the ground state of a given Hamiltonian and the two states $|\psi_1\lb$ and $|\psi_2\lb$ are excited states.

We can explicitly write the two states as ($\ep_{1,2}$ are infinitesimally small parameters)
\ba
&& |\psi_1\lb=\s{1-|\ep_1|^2}|\psi_0\lb+\ep_1|\alpha\lb, 
\ \ \ \  |\psi_2\lb=\s{1-|\ep_2|^2}|\psi_0\lb+\ep_2|\beta\lb, 
\ea
where we assume $\la\alpha|\psi_0\lb=\la\beta|\psi_0\lb=0$ and the unit norm 
$\la\alpha|\alpha\lb=\la\beta|\beta\lb=1$.

We would like to consider the sign of the difference:
\ba
S(\tau^{1|2}_A)+S(\tau^{2|1}_A)
-S(\rho_{A}^1)-S(\rho_{A}^2), \label{difr}
\ea
up to $O(\ep^2)$, where $\rho_{A}^1=\mbox{Tr}_B[|\psi_1\lb\la\psi_1|]$ and  $\rho_{A}^2=\mbox{Tr}_B[|\psi_2\lb\la\psi_2|]$.
Note that when $S(\tau^{1|2}_A)$ is real valued, which we assume in the main context of this paper,
(\ref{difr}) is identical to the twice of the difference 
\ba
S(\tau^{1|2}_A)-\frac{1}{2}\left(S(\rho_{A}^1)+S(\rho_{A}^2)\right). \label{difpe}
\ea

The transition matrix deviates from $\rho^0_A=\mbox{Tr}_B[|\psi_0\lb\la \psi_0|]$ as  
\ba
\tau^{1|2}_A\simeq \rho_A+\ep_1\mbox{Tr}[|\alpha\lb\la \psi_0|]+\ep^*_2\mbox{Tr}[|\psi_0\lb\la \beta|]+O(\ep^2),\nonumber
\ea
where we noted $\la\psi_2|\psi_1\lb\simeq 1+O(\ep^2)$.
By using (\ref{perty}) repeatedly, this leads to (up to $O(\ep^2)$)
\ba
&& S(\tau^{1|2}_A)+S(\tau^{2|1}_A)-S(\rho_{A}^1)-S(\rho_{A}^2)\no
&& \simeq \frac{\la\psi_1|H_A|\psi_2\lb}{\la \psi_1|\psi_2\lb}+ \frac{\la\psi_2|H_A|\psi_1\lb}{\la \psi_2|\psi_1\lb}-\la\psi_1|H_A|\psi_1\lb-\la\psi_2|H_A|\psi_2\lb+\Delta^{(2)}S, \no
&& \simeq -\left(\ep^*_1\la\alpha|-\ep^*_2\la\beta|\right)|H_A|
\left(\ep_1|\alpha\lb-\ep_2|\beta\lb\right)+\Delta^{(2)}S,\nonumber
\ea
where the linear $O(\ep)$ terms do cancel.
Here $\Delta^{(2)}S$ is the quadratic contribution from the last integral term in (\ref{pertgh}).

In particular, if we consider the special perturbation where $|\alpha\lb=|\beta\lb$ and then we find
\ba
&& S(\tau^{1|2}_A)+S(\tau^{2|1}_A)-S(\rho_{A}^1)-S(\rho_{A}^2) \no
&& \simeq -|\ep_1-\ep_2|^2\cdot \la\alpha|H_A|\alpha\lb+|\ep_1-\ep_2|^2\!
\int^\infty_0 \!dt \mbox{Tr}\left[\frac{t}{(t+\rho_A)^2}
\mbox{Tr}_B\left[|\alpha\lb\la \psi_0|\right] \frac{1}{t+\rho_A}
\mbox{Tr}_B\left[|\psi_0\lb \la\alpha|\right]
\right]\no
&&+|\ep_1-\ep_2|^2\!
\int^\infty_0\! dt \mbox{Tr}\left[\frac{t}{(t+\rho_A)^2}\mbox{Tr}_B\left[|\psi_0\lb \la\alpha|\right]
 \frac{1}{t+\rho_A}
\mbox{Tr}_B\left[|\alpha\lb\la \psi_0|\right]
\right]. \nonumber
\ea
This shows that the above difference is proportional to $|\ep_1-\ep_2|^2$. However the sign of the 
quadratic is not definite from the above analysis. Indeed we will find that it can be both negative and positive 
below. Nevertheless as we will see in appendix F and the main context of this article, the sign turns out to be non-positive for quantum field theories.

\subsection{Perturbations in Two Qubit System}

For the two qubit system, we choose 
\ba
&& |\psi_1\lb_{AB}=\cos\theta_1|00\lb_{AB}+\sin\theta_1|11\lb_{AB},
\ \ \ \  |\psi_2\lb_{AB}=\cos\theta_2|00\lb_{AB}+\sin\theta_2|11\lb_{AB},\nonumber
\ea
where we assume $0\leq \theta_1,\theta_2\leq \frac{\pi}{2}$.
The pseudo entropy is computed as \cite{Nakata:2020fjg}
\ba
&& S(\tau^{1|2}_A) =-\left(\frac{\cos\theta_1\cos\theta_2}{\cos(\theta_1-\theta_2)}\right)\cdot \log\left(\frac{\cos\theta_1\cos\theta_2}{\cos(\theta_1-\theta_2)}\right)-\left(\frac{\sin\theta_1\sin\theta_2}{\cos(\theta_1-\theta_2)}\right)\cdot \log\left(\frac{\sin\theta_1\sin\theta_2}{\cos(\theta_1-\theta_2)}\right).\nonumber
\ea

\begin{figure}
\begin{center}
\includegraphics[scale=0.4]{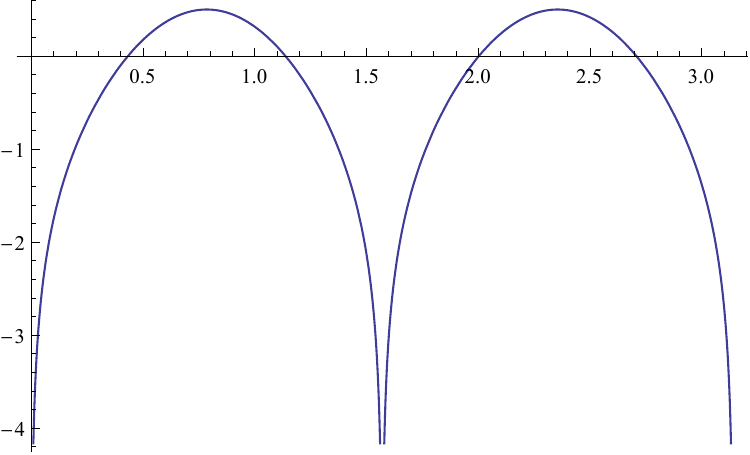}
\end{center}
\caption{The plot of the difference $q(\theta)$ in  (\ref{fthea}).}
\label{fig:eqqvs}
\end{figure}

We are interested in a small perturbation $\theta_2-\theta_1=\delta \ll 1$. Then the interesting difference looks like
\ba
2S(\tau^{1|2}_A) -S(\rho^1_A)-S(\rho^2_A)
\simeq q(\theta_1)\delta^2+O(\delta^3),\nonumber
\ea
where we find
\ba
q(\theta)=\frac{1}{2}+\frac{\cos 2x}{2}\cdot \log \tan^2\theta.
\label{fthea}
\ea
This function is plotted in Fig.\ref{fig:eqqvs}. It is not always negative. In particular when the state $|\psi_1\lb$ is highly entangled, the difference tends to be positive.

\section{Appendix F: Pseudo Entropy for Perturbed CFTs}

Here we analyze the change of the pseudo entropy $S(\tau^{1|2}_A)$ when we perturb a CFT vacuum 
by a primary operator in two dimension. We choose $|\psi_1\lb$ to be the original CFT vacuum and 
$|\psi_2\lb$ to be the vacuum in the perturbed theory. We will calculate this both 
from the field theoretic and holographic approaches.

\subsection{CFT Perturbations}

Consider a two dimensional CFT perturbed by a primary operator $O(x)$ with the (chiral) conformal 
dimension $h$:
\ba
S=S_{CFT}+\int d^2x \lambda(x)O(x).  \label{pertl}
\ea
To describe a transition matrix, we assume 
\ba
\lambda(x)=\lambda \cdot \theta(x_1),
\ea
where $x_1$ is the coordinate of the Euclidean time. Note that this is chosen such that 
the initial state is the original CFT vacuum, while the final state is the ground state of the 
perturbed theory (\ref{pertl}). 

We introduce the complex coordinate $(w,\bar{w})$ such that $w=x_2+ix_1$ and choose the subsystem $A$
to be $0\leq x_2\leq l$ at the time $x_1=0$.
In this setup,  we have
\ba
\mbox{Tr}[(\tau^{1|2}_A)^n]=\frac{\la e^{-\int d^2w \lambda(w,\bar{x}) O(w,\bar{w})}\lb_{\Sigma_n}}
{\left(\la e^{-\int d^2w  \lambda(w,\bar{x}) O(w,\bar{w})}\lb_{\Sigma_1}\right)^n}.
\ea
Here $\Sigma_n$ is the $n$-sheeted Riemann surface 
obtained by gluing $n$ complex planes along the cut $A$. 

The reduced transition matrix at $\lambda=0$ coincides with the reduced density matrix for the CFT vacuum $\rho^1_A$. Thus by using the fact that the one-point function in a CFT vanishes and by expanding up to the quadratic order we have
\ba
 \frac{\mbox{Tr}[(\tau^{1|2}_A)^n]}{\mbox{Tr}[(\rho^1_A)^n]}
 && \simeq 1+\frac{\lambda^2}{2}\int_{\Sigma^{+}_n} dw^2_1 dw^2_2 \la O(w_1)O(w_2)\lb_{\Sigma_n}
-\frac{n\lambda^2}{2}\int_{\Sigma^{+}_1} dw^2_1 dw^2_2 \la O(w_1)O(w_2)\lb_{\Sigma_1},\label{intper}
\ea
where $\Sigma^{+}_n$ denotes the upper half of the surfaces $\Sigma_n$, 
where the perturbation is restricted.
The difference between the $n$-th Renyi pseudo entropy and the original $n$-th Renyi entropy is 
\ba
S^{(n)}(\tau^{1|2}_A)-S^{(n)}(\rho_A)=\frac{1}{1-n}\log 
\left[\frac{\mbox{Tr}[(\tau_A)^n]}{\mbox{Tr}[(\rho_A)^n]}\right].
\ea
By taking the limit $n\to 1$, we obtain the difference $S(\tau^{1|2}_A)-S(\rho_A^1)$
between the pseudo entropy and entanglement entropy.

The two point function on a complex plane $\Sigma_1$ reads 
\ba
\la O(w_1)O(w_2)\lb_{\Sigma_1}=\frac{1}{|w_1-w_2|^{4h}}.
\ea
To calculate the two point functions on $\Sigma_n$, we perform the conformal map from 
$\Sigma_n$ into a complex plane R$^2$:
\ba
z^n=\frac{w}{w-l}.  \label{mapn}
\ea
This gives
\ba
\int_{\Sigma^{+}_n} d^2w_1 d^2w_2 \la O(w_1)O(w_2)\lb_{\Sigma_n}  &=& \int_{P^{+}_n}d^2z_1 d^2z_2 \left|\frac{dz_1}{dw_1}\right|^{2(1-h)}
\left|\frac{dz_2}{dw_2}\right|^{2(1-h)} \frac{1}{|z_1-z_2|^{4h}}\no
&=&\int_{P^{+}_n}\!\!d^2z_1 d^2z_2  \left|\frac{nLz_1^{n-1}}{(z_1^n-1)^2}\right|^{2(1-h)}
\left|\frac{nLz_2^{n-1}}{(z_2^n-1)^2}\right|^{2(1-h)}\!\frac{1}{|z_1-z_2|^{4h}},\nonumber
\ea
where $P^{+}_n$ is the image of $\Sigma^+_{n}$ by the conformal map (\ref{mapn}).
Explicitly we have 
\ba
P^+_n=\{z=r e^{i\theta}|0\leq r<\infty,\ \theta\in Q_n\},
\ea
where
\ba
Q_n=\cup_{k=0}^{n-1}\left[\frac{2\pi k}{n},\frac{2\pi k}{n}+\frac{\pi}{n}\right].  \label{qset}
\ea
We call $n$ disconnected regions in $Q_n$ as $n$ chambers.

In the actual computations, we need a UV regularization when $z_1$ and $z_2$ get closer.
To have a universal treatment of such a cut off,  we rewrite the latter $w$-integral in
(\ref{intper}) in terms of $z$ coordinate as follows
\ba
\int_{\Sigma^{+}_1} dw^2_1 dw^2_2 \la O(w_1)O(w_2)\lb_{\Sigma_1}&=&\int_{P^{+}_1}d^2z_1 d^2z_2 \left|\frac{nLz_1^{n-1}}{(z_1^n-1)^2}\right|^{2}
 \left|\frac{nLz_2^{n-1}}{(z_2^n-1)^2}\right|^{2} \frac{1}{|w_1-w_2|^{4h}}\no
&=&\int_{P^{+}_1}\!\!d^2z_1 d^2z_2  
\left|\frac{nLz_1^{n-1}}{(z_1^n-1)^2}\cdot\frac{nLz_2^{n-1}}{(z_2^n-1)^2}\right|^{2(1-h)}
\!\frac{ |G(z_1,z_2)|^{4h}}{|z_1-z_2|^{4h}},\nonumber
\ea
where we introduced
\ba
G(z_1,z_2)=n\frac{(z_1-z_2)(z_1z_2)^{\frac{n-1}{2}}}{z^n_2-z^n_1}.
\ea
The region $P^+_1$ is defined by 
\ba
P^+_1=\{z=r e^{i\theta}|0\leq r<\infty,\ 0\leq \theta\leq \frac{\pi}{n}\},
\ea
i.e. $\frac{1}{n}$ fraction of  $P^{+}_n$ or  a single chamber.

It is useful to note 
\ba
\lim_{z_2\to z_1}G(z_1,z_2)=1,\ \   \mbox{and} \ \ \ |G(z_1,z_2)|\leq 1\ \  \label{dwww}
\ea

Then we can evaluate as follows 

\ba
&& \log \left[\frac{\mbox{Tr}[(\tau^{1|2}_A)^n]}{\mbox{Tr}[(\rho^1_A)^n]}\right] \no
&& \simeq \frac{\lambda^2}{2}\int_{P^{+}_n}d^2z_1 d^2z_2  |v(z_1)v(z_2)|^{2(1-h)}\frac{1}{|z_1-z_2|^{4h}}
\!-\frac{n\lambda^2}{2}\!\!\int_{P^{+}_1}\!\!d^2z_1 d^2z_2   |v(z_1)v(z_2)|^{2(1-h)}\!
\left|\frac{G(z_1,z_2)}{z_1-z_2}\right|^{4h},
\label{explt}
\ea 
where
\ba
v(z)=\frac{nLz^{n-1}}{(z^n-1)^2}.
\ea
It is clear from $G \leq 1$  (\ref{dwww}) that this difference is positive 
\ba
 \log \left[\frac{\mbox{Tr}[(\tau^{1|2}_A)^n]}{\mbox{Tr}[(\rho^1_A)^n]}\right] \geq 0,
\ea 
which gives the non-positivity of the difference
\ba
S^{(n)}(\tau^{1|2}_A)-S^{(n)}(\rho^1_A)\leq 0.
\ea
This is because the difference is bounded from below by setting $G=1$. If we set $G=1$,
the second term in (\ref{explt}) 
is canceled by the contributions from the first term where $z_1$ and $z_2$ are 
in the same chamber. Therefore totally the contributions from the first term where $z_1$ and $z_2$ are in different chambers remain, which are clearly positive.

\subsection{Exact Marginal Perturbation $h=1$}

Let us estimate the leading divergent contribution for
the exactly marginal perturbation $h=1$. Since such a divergence arises when $z_1\simeq z_2$,
we set $G\simeq 1$ and obtain
\ba
&& \log \left[\frac{\mbox{Tr}[(\tau^{1|2}_A)^n]}{\mbox{Tr}[(\rho^1_A)^n]}\right] 
\simeq \frac{\lambda^2}{2}\int_{P^{+}_n}d^2z_1 d^2z_2  \frac{1}{|z_1-z_2|^{4}}
-\frac{n\lambda^2}{2}\int_{P^{+}_1}d^2z_1 d^2z_2  \frac{1}{|z_1-z_2|^{4}}.\nonumber
\ea 
The divergences when $z_1\simeq z_2$ are canceled out when both $z_1$ and $z_2$ are 
in the same chamber. Thus the leading logarithmic divergence arises where
 $z_1$ and $z_2$ are in the different chambers. In this case the divergence occurs in the two 
limits $z_1,z_2\to 0$ or $z_1,z_2\to \infty$ corresponds to the limits that the coordinate 
$w_1$ and $w_2$ both get closer to either of the two end points of the interval $A$.

This consideration leads to the following estimation
\ba
&& \log \left[\frac{\mbox{Tr}[(\tau^{1|2}_A)^n]}{\mbox{Tr}[(\rho^1_A)^n]}\right] 
\simeq \frac{n\lambda^2}{2}\!\int^{\infty}_0\! dr_1\! \int^\infty_0\! dr_2 
\int^{\frac{\pi}{n}}_0\! d\theta_1\! \sum^{n-1}_{k=1}\int^{\frac{2\pi k}{n}+\frac{\pi}{n}}_{\frac{2\pi k}{n}}\!
d\theta_2\frac{r_1r_2}{|r_1e^{i\theta_1}-r_2e^{i\theta_2}|^4}.\nonumber
\ea

We write $(r_1,r_2)=\rho (\cos\phi,\sin\phi)$, where $0\leq \rho<\infty$ and 
$0\leq\phi\leq\frac{\pi}{2}$. We can regulate the divergent at $\rho=0$ 
by setting $\delta<\rho$, where
 the UV cut off $\delta$ is expresses in terms of the lattice spacing $\ep$ as follows:
\ba
\delta\sim \frac{\ep}{L}.
\ea
A similar regularization for $\rho\to \infty$ gives the same contribution. Totally we obtain the doubled 
contribution given by
\ba
 \log \left[\frac{\mbox{Tr}[(\tau^{1|2}_A)^n]}{\mbox{Tr}[(\rho^1_A)^n]}\right]
\simeq n\lambda^2\cdot c_n\cdot \log\delta^{-1},
\ea
where we defined
\ba
c_n\!=\!\int^{\frac{\pi}{2}}_0\! d\phi \!\int^{\frac{\pi}{n}}_{0}\!d\theta_1 \!
 \sum^{n-1}_{k=1}\!\int^{\frac{2\pi k}{n}+\frac{\pi}{n}}_{\frac{2\pi k}{n}}\!d\theta_2\frac{1}
{\left(1-\sin2\phi\cos(\theta_1-\theta_2)\right)^2}.\nonumber
\ea

Thus we get 
\ba
S^{(n)}(\tau^{1|2}_A)-S^{(n)}(\rho^1_A)\simeq \frac{nc_n}{1-n}\lambda^2 \cdot \log \frac{l}{\ep}.
\ea
Since we can confirm that $c_n$ is positive and monotonically increasing function of $n$,
the above difference is negative in the limit $n\to 1$ and thus we obtain
$S(\tau^{1|2}_A)-S(\rho^1_A)< 0$.  Since the exact marginal perturbation does not change the central 
charge $c$, the logarithmic divergence in $S(\rho^1_A)$ has the same coefficient as that of $S(\rho^2_A)$.
Thus we can conclude that  $\Delta S_{12}=S(\tau^{1|2}_A)-S(\rho^1_A)/2-S(\rho^2_A)/2<0$ under an exactly marginal perturbation. In other words, we have the expression in (\ref{stauc}), where
the function $f(\lambda)$ behaves as $f(\lambda)=1+g\lambda^2+O(\lambda^3)$. 
The coefficient $g$ is negative because $\frac{c}{3}g=-\frac{dc_n}{dn}|_{n=1}$.

\subsection{Holographic Analysis}
Janus solutions 
\cite{Bak:2003jk,Freedman:2003ax,Clark:2004sb,Clark:2005te,DHoker:2006vfr}
provides us with a full order answer to the exactly marginal perturbation of a CFT in any dimension.
Below we would like to evaluate the holographic pseudo entropy from the minimal areas in 
Janus solutions.

The $d+1$ dimensional Janus solutions take the general form: 
\ba
ds^2=d\rho^2+e^{h(\rho)}ds^2_{AdS(d)},  \label{xxe}
\ea
where $ds^2_{AdS(d)}$ is $d$ dimensional AdS metric
\ba
ds^2_{AdS(d)}=\frac{dy^2+\sum_{\mu=1}^{d-1}dx^i dx_i}{y^2}.
\ea
We are interesting in the holographic pseudo entropy at the time slice of the dual CFT 
defined by $\rho=y=0$. The subsystem $A$ sits on this $d-1$ dimensional time slice.

We assume the Z$_2$ invariance $h(\rho)=h(-\rho)$ so that the minimal surface $\Gamma_A$ sits on the slice $\rho=0$.  Also we assume both the future and past infinity describe two different CFT vacua 
$|\psi_1\lb$ and $|\psi_2\lb$ with the same central charge $c$.
This requires $h(\rho)\simeq \pm 2\rho$ in the limit $\rho\to\pm\infty$. 
The coordinate $x$ is the space direction of the dual interface CFT. The Euclidean time direction of the CFT
is $(\rho,y)$ direction as usual.  $\rho=\infty$ (and $\rho=-\infty$) corresponds to the upper (and lower) half plane of the interface CFT. 

When $d=2$, we choose the subsystem $A$ as an interval $-l/2\leq x \leq l/2$ at $t=0$ (i.e. the location of Janus interface) and calculate its holographic pseudo entropy. Due to the Z$_2$ symmetry,  $\Gamma_A$ is the geodesic on the $\rho=0$ slice. If we write the cut off of $y$ as $\ti{\ep}$ we have the following estimation of holographic pseudo entropy:
\ba
S(\tau^{1|2}_A)=\frac{c}{3}e^{\frac{h(0)}{2}}\log\frac{l}{\ti{\ep}},
\ea
where $c$ is the central charge for the CFT.

Below we would like to work out whether this pseudo entropy is smaller than the original CFT entanglement 
entropy:
\ba
S_A(|\psi_1\lb)=S_A(|\psi_2\lb)=\frac{c}{3}\log\frac{l}{\ep}
\ea
We do not care about the difference between $\ti{\ep}$ and $\ep$ as this only leads to a subleading
difference.  To argue the difference (\ref{difpe}) is non-positive, we need to confirm
\ba
h(0)\leq 0.  \label{inrewdf}
\ea
We can generalize this to any higher  $d$ dimensions straightforwardly  and we can easily confirm that
the difference is non-positive if when (\ref{inrewdf}) is satisfied.

Below we would like to argue,   (\ref{inrewdf}) is always true for any (physically sensible) Janus solutions
in any dimensions. 
For this we would like to first impose an Euclidean version of null energy condition
\ba
R_{\mu\nu}N^\mu N^\nu \geq 0,
\ea
where $N^\mu$ is arbitrary null vector. In our Euclidean setup (\ref{xxe}),  we choose
\ba
N_{(1)}^\mu=(0,1,i),\ \ \ N_{(2)}^\mu =(1,e^{-h(\rho)}y i,0).
\ea
   The first one $N_{(1)}$ is trivial $R_{\mu\nu}N^\mu N^\nu=0$ but the second one leads to 
\ba
2e^{-h}-h''\geq 0.  \label{ineqrf}
\ea

In an explicit example of 3D Janus solution of Einstein-dilaton theory \cite{Bak:2007jm}, we have 
\ba
e^{2h(\rho)}=\frac{1}{2}+\frac{\s{1-2\gamma^2}}{2}\cosh(2\rho).
\ea
In this case we find 
\ba
2e^{-h}-h''=2\gamma^2 e^{-h},
\ea
which is indeed positive.

The Z$_2$ symmetry and asymptotic behavior assumptions are
\ba
&& h'(0)=0, \ \ \ h(\rho)=h(-\rho),\ \ \  h(\rho)\simeq  2\rho\ \ \ (\rho\to \infty). 
\ea

We assume $h'(\rho)\geq 0$ as is true in known Janus solutions \cite{Bak:2003jk,Freedman:2003ax,Clark:2004sb,Clark:2005te,DHoker:2006vfr}.
Then we can multiply $h'$ with (\ref{ineqrf}) and get
\ba
\frac{1}{2}\de_\rho(h'^2)\leq \de_\rho[-2e^{-h}].
\ea
By integrating this from $\rho=0$ to $\rho=\infty$, we find 
\ba
2\leq 2 e^{-h(0)},
\ea
which leads to the expected inequality (\ref{inrewdf}).

\end{document}